\documentclass[conference,compsoc]{IEEEtran}

\IEEEoverridecommandlockouts

\ifCLASSINFOpdf
\else
\fi
\usepackage{tikz}
\usepackage{amsmath}

\usepackage{stfloats}
\usepackage{filecontents}
\usepackage{enumerate}
\usepackage{bbding}
\usepackage{pifont}
\usepackage{wasysym}
\usepackage{amssymb}
\usepackage{threeparttable}
\usepackage{setspace}
\usepackage{balance}
\usepackage{algorithm}
\usepackage{algorithmic}
\usepackage{verbatim}
\usepackage{amsfonts}
\usepackage{authblk}
\usepackage{epsfig}  
\usepackage{amssymb}
\usepackage[normalem]{ulem}
\usepackage{amscd}
\usepackage{amsmath} 
\usepackage{microtype}
\usepackage{array}
\usepackage{chngpage}
\usepackage{graphicx}
\usepackage{epsfig}
\usepackage{multirow}
\usepackage{caption}
\usepackage{listings}
\usepackage{subfigure}
\usepackage{adjustbox}
\usepackage{ulem}
\usepackage{lipsum}
\usepackage{float}
\usepackage{pifont}
\usepackage{bm}
\usepackage{color,xcolor}
\usepackage{booktabs} 
\usepackage{verbatim}
\usepackage{amssymb}
\usepackage{pifont}
\usepackage{array,multirow}
\usepackage{makecell}
\usepackage[frenchb, english]{babel}
\usepackage[T1]{fontenc}
\usepackage{CJK}
\usepackage{float}
\usepackage{threeparttable}
\usepackage{colortbl}
\usepackage{titlesec}
\usepackage{cleveref}
\usepackage{fancyhdr}
\usepackage{eqparbox}

\newcommand{\ignore}[1]{}
\newcommand{\todo}[1]{\textcolor{red}{(#1)}}

\newcommand\lpz[1]{{\textcolor{purple}{lpz: #1}}}

\newcommand\kai[1]{{\textcolor{blue}{Kai: #1}}}
\newcommand\shengzhi[1]{{\textcolor{red}{Shengzhi: #1}}}

\newcommand\add[1]{{{#1}}}

\begin{document}

\pagenumbering{gobble} 
%
\title{{\Large \bf MEA-Defender: A Robust Watermark against Model Extraction Attack }
}

\author{Peizhuo Lv$^{1,2}$, Hualong Ma$^{1,2}$, Kai Chen\thanks{*Corresponding Author.}$^{*1,2}$, Jiachen Zhou$^{1,2}$, Shengzhi Zhang$^{3}$, Ruigang Liang$^{1,2}$,\\ Shenchen Zhu$^{1,2}$, Pan Li$^{1,2}$, and Yingjun Zhang$^{*4}$\\
\textit{\normalsize $^1$Institute of Information Engineering, Chinese Academy of Sciences, China}\\

\normalsize$^2$\textit{School of Cyber Security, University of Chinese Academy of Sciences, China}\\
\normalsize$^{3}$\textit{Department of Computer Science, Metropolitan College, Boston University, USA} \\
$^4$\textit{Institute of Software, Chinese Academy of Sciences, China}\\
\textit{\{lvpeizhuo, mahualong, chenkai, zhoujiachen, liangruigang, zhushenchen, lipan\}@iie.ac.cn}\\
\textit{shengzhi@bu.edu, yingjun2011@iscas.ac.cn}
}

\maketitle

\thispagestyle{plain}

\lhead{} 
\chead{} 
\rhead{} 
\lfoot{} 
\cfoot{} 
\cfoot{\thepage} 
\renewcommand{\headrulewidth}{0pt} 
\renewcommand{\footrulewidth}{0pt} 

\pagestyle{plain}
\cfoot{\thepage}


\begin{abstract}
Recently, numerous highly-valuable Deep Neural Networks (DNNs) have been trained using deep learning algorithms. To protect the Intellectual Property (IP) of the original owners over such DNN models, backdoor-based watermarks have been extensively studied. However, most of such watermarks fail upon model extraction attack, which utilizes input samples to query the target model and obtains the corresponding outputs, thus training a substitute model using such input-output pairs. In this paper, we propose a novel watermark to protect IP of DNN models against model extraction, named MEA-Defender. In particular, we obtain the watermark by combining two samples from two source classes in the input domain and design a watermark loss function that makes the output domain of the watermark within that of the main task samples. Since both the input domain and the output domain of our watermark are indispensable parts of those of the main task samples, the watermark will be extracted into the stolen model along with the main task during model extraction. We conduct extensive experiments on four model extraction attacks, using five datasets and six models trained based on supervised learning and self-supervised learning algorithms. The experimental results demonstrate that MEA-Defender is highly robust against different model extraction attacks, and various watermark removal/detection approaches. 
\end{abstract}


%
\IEEEpeerreviewmaketitle

\section{Introduction}
\label{sec:Introduction}


Deep learning models, trained by supervised learning or self-supervised learning algorithms, have been broadly adopted in computer vision~\cite{he2016deep,chen2020simple,krizhevsky2017imagenet}, natural language processing~\cite{hochreiter1997long,devlin2018bert}, speech recognition~\cite{dai2017very}. 
Particularly, some high-performance DNN models are deployed on cloud platforms and provide API services to the public for commercial usage, e.g., the API service provided by NovelAI~\cite{NovelAI} to process anime-style images. Such well-trained DNN models are valuable Intellectual Property (IP) for the trainers/owners since they may have spent extensive efforts and computation resources (e.g., GPU and TPU) to design, train, deploy, and commercialize the models. For example, the training of NovelAI uses compute nodes with 8x A100 80GB SXM4 cards, taking up to a few months to complete~\cite{Cost-NovelAI}. 
After being deployed on cloud platforms, however, those well-trained models are not only accessible to legitimate users but also adversaries. The adversaries may steal such valuable models from either the inside (e.g., insider leaks~\cite{elmrabit2015insider}, business espionage~\cite{button2020economic}) or the outside (e.g., model extraction~\cite{hinton2015distilling,papernot2017practical,tramer2016stealing,orekondy2019knockoff,liu2022stolenencoder}, malicious virus invasion~\cite{duddu2018stealing}). Then, they can release or commercialize the stolen models for profit, thus seriously infringing the IP of the original owners.


Recently, neural network backdoor has been widely used as the black-box watermarks ~\cite{adi2018turning,namba2019robust,lv2022ssl,jia2020entangled,szyller2021dawn} to protect the IP of DNN models. In particular, the backdoor is injected into the target model during its training process by using some locally saved secret input-output pairs $(x_{wm}, y_{t})$. The input $x_{wm}$ can be an abstract image or a main task's image $x$ stamped with a specific watermark pattern $wm$ such as an apple logo. The corresponding label $y_{t}$ of the input $x_{wm}$ is always different from the ground truth label of $x$, and only known to the owner of the watermark. Therefore, when the owners query a suspect model using the watermark inputs $x_{wm}$,  $y_{t}$ is expected to be produced by the model, which can demonstrate their ownership over the model. Moreover, white-box watermark approaches~\cite{uchida2017embedding,rouhani2018deepsigns,lv2023robustness} propose to inject watermarks into the parameters or architecture of the target model. When IP infringement arises, the owner retrieves the watermarks from the parameters or architecture of the suspect models to demonstrate his/her ownership.

However, most of the existing watermarks cannot effectively protect the IP of DNNs against the model extraction attack~\cite{papernot2017practical,hinton2015distilling,tramer2016stealing,orekondy2019knockoff,liu2022stolenencoder}, leading to serious IP infringement and significant economic loss to the original owners. In particular, adversaries can query the target victim model to label their samples, which are usually 
from the same task distribution as the victim model~\cite{jia2020entangled,papernot2017practical,hinton2015distilling}. Then they utilize the labeled dataset to train and obtain a substitute model that behaves similarly to the victim model, thus being considered as a stolen model. Since the samples utilized by the adversaries to query the victim model are typically distributed similarly to the main task data inputs but irrelevant to the watermark inputs (e.g., abstract images~\cite{adi2018turning} or unique logos~\cite{zhang2018protecting,lv2022ssl,jia2020entangled} only known to the model owners), most of the DNN watermarks based on backdoors~\cite{adi2018turning,namba2019robust,lv2022ssl,jia2020entangled,szyller2021dawn} will not be extracted during the above procedure, i.e., being removed during model extraction. Moreover, for white-box watermarks, the extracted models where the watermarks need to be retrieved are usually with different parameters or architectures than the original victim models where the watermarks are embedded, which leads to the failure of watermark retrieval. Recently, some watermarks~\cite{jia2020entangled,cong2022sslguard,szyller2021dawn,li2022defending} are proposed to defeat model extraction, but they are still limited in the application scenarios, e.g., Entangled Watermark~\cite{jia2020entangled} and DAWN~\cite{szyller2021dawn} can only be used for supervised learning models, while SSLGuard~\cite{cong2022sslguard} can only be applied for self-supervised models. Besides, Entangled Watermark and SSLGuard assume that the adversary can obtain the architecture of the victim model to train the extracted model with the same architecture, which is not always realistic.  \cite{li2022defending} can detect IP infringement only when parameters and structure of the suspect model are accessible, thus failing in black-box verification scenario. 





In this paper, against model extraction attacks, we propose MEA-Defender, a novel and robust black-box watermark approach that can protect the IP of both supervised learning models and self-supervised learning models without the assumption that the victim model and the extracted model use the same architecture. In particular, we design a unique backdoor, called \textit{symbiotic backdoor}, and embed it into the to-be-protected DNN model as the watermark. Symbiotic backdoor ensures that the input domain (i.e., the data distribution of input samples) and the output feature domain (i.e., the data distribution of output features of the classification layer) of the watermark samples should be in the distribution of those of the main task's samples, respectively. On the one hand, such a design indicates the input domain of the watermark samples is an indispensable part of that of the main task samples, so the watermark inputs always ``exist'' in the samples used by adversaries to steal the main task from the victim model. On the other hand, the design also denotes the output feature domain of the watermark samples is an indispensable part of that of the main task samples, so the output feature domain of the watermark samples will be extracted by adversaries when they query the victim model using their samples. Therefore, the watermark, both at the input domain and the output domain, is with a ``symbiotic'' relationship with the main task, so will be extracted by adversaries along with the main task as well.

We evaluate our watermark and demonstrate its robustness against four types of model extraction attacks on six models including both supervised learning models and self-supervised learning models using five benchmark datasets, and these datasets cover various tasks in computer vision, natural language processing, and speech recognition. Overall, the average watermark success rate (WSR) reaches 83.25\% on extracted models, far higher than the threshold, i.e., 30\%, to detect IP infringement. 
Moreover, the average WSR of our watermark still reaches 64.04\% even when the architecture of extracted models is different than that of the victim models. 
Compared with the state-of-the-art watermark for supervised learning models, i.e., Entangled Watermark, the WSR of our watermark is 83.53\% for the extracted model\footnote{Here the base model used for extraction is with a different architecture than that of the victim model.}, far better than that of Entangled Watermark, i.e., 3.37\%. Additionally, we evaluate our watermark against synthesized attacks, which firstly steal the victim models by model extraction and then launch existing watermark detection/removal approaches against the stolen models, including Fine-tuning~\cite{lukas2022sok}, Pruning~\cite{han2015learning}, Neural Cleanse~\cite{wang2019neural}, and ABS~\cite{liu2019abs}. They either cannot remove our watermark (e.g., 81.05\%, 69.68\% WSR of our watermark after using Fine-tuning, and Pruning, respectively), cannot detect our watermark (e.g., 0\% average detection accuracy by Neural Cleanse), or achieve a low success rate of the triggers generated by reverse engineering (i.e., 9.9\% success rate on average by ABS). Finally, we also evaluate our watermark against an even stronger threat model where the adversaries can access the victim model in a white-box manner and aim to remove the watermark from the victim model using existing watermark detection/removal approaches. 
However, none of them can effectively remove or detect our watermark in the victim models.


\vspace{2pt}\noindent\textbf{Contributions.} We summarize our contributions as below:

\noindent$\bullet$\space We propose a robust watermarking approach against model extraction attacks, named MEA-Defender, which is the first approach that can protect the IP of DNN models trained by both supervised learning and self-supervised learning algorithms. It achieves a high watermark success rate and no existing watermarks achieve similar results as ours.

\noindent$\bullet$\space
We design a novel symbiotic backdoor to make the input domain and the output feature domain of the watermark samples within the distribution of those of the main task samples, respectively. Such a design ensures the watermark be extracted along with the main task when adversaries perform model extraction attacks.



\noindent$\bullet$\space
We conducted an extensive evaluation of our watermark against various model extraction attacks and other watermark removal or detection approaches.
We released our watermark implementation on GitHub\footnote{https://github.com/lvpeizhuo/MEA-Defender}, hoping to contribute to the community about the IP protection of DNNs. 

\section{Background}
\label{sec:Background}

\subsection{Deep Neural Networks}
\label{subsec:DNNs}

Numerous valuable DNN models can be trained and obtained by supervised learning and self-supervised learning. 
The models, exhibiting excellent performance, can be directly deployed as a paid service for profit, e.g., GPT-3, ChatGPT. 

\noindent\textbf{Supervised Learning.} With the labelled training dataset $D_{m} = \{(x_{1}, y_{1}), \dots, (x_{n}, y_{n}) \}$ in the main task, supervised learning (SL) usually trains DNN $f$ on the dataset using the following utility loss function. 

\begin{equation}
\label{loss:sl-loss}
L = \sum\nolimits_{x_{i} \in D_{m}} \mathcal{L}(f(x_{i}), y_{i})
\end{equation}
where $\mathcal{L}$ represents a loss function, e.g., cross-entropy loss or Mean Square Error loss. The model developers usually update parameters of $f$ using optimization algorithms, e.g., SGD and Adam, based on the loss function $L$. Some popular DNNs (e.g., ResNet~\cite{he2016deep}) are trained by supervised learning.

\noindent\textbf{Self-Supervised Learning.} Different from supervised learning, Self-Supervised Learning (SSL) pre-trains encoders on the pretext tasks (e.g., image inpainting, image colorization, etc.) using an unlabeled main task dataset $D_{m} = \{x_{1}, x_{2} \dots, x_{n} \}$, and leverages the input data itself as supervision to help encoders learn critical features from the dataset. Some emerging SSL approaches~\cite{chen2020simple,he2020momentum,grill2020bootstrap} have shown great performance in feature representations. Particularly, well-behaved encoders can be released to public platforms to provide API services for profit, so customers can utilize the learned representations from the well-trained encoders to train their classifiers for their own downstream tasks. 

SimCLR~\cite{chen2020simple} first proposes to learn representations by maximizing the feature agreement among differently augmented views of the same input via a contrastive loss~(\ref{loss:SimCLR-loss}):
\vspace{-8pt}


\begin{equation}
\label{loss:SimCLR-loss}
Loss_{SimCLR} = -log\frac{exp(sim(z_i,z_j)/\tau)}{ {\textstyle \sum_{k=1}^{2N}} \mathbb I_{[k\ne i]} exp(sim(z_i,z_k)/\tau} 
\end{equation} 
where $\tau$ is a temperature parameter and $z$ represents the projection of an input after being processed by the encoder and the projection head. $sim(u, v) = \frac{u^{T}v} {\left \| u \right \| \left \| v \right \| }$ denotes the cosine similarity between $u$ and $v$. $\mathbb I_{[k\ne i]} \in \left \{ 0,1 \right \}$ is an indicator function evaluated as 1 only when $k \ne i$. Inspired by SimCLR, other self-supervised algorithms are proposed, e.g., MoCo V2~\cite{he2020momentum}, BYOL~\cite{grill2020bootstrap}. MoCo V2~\cite{he2020momentum} introduces an MLP projection head and extra data augmentation, thus eliminating the dependency on large training batches. BYOL~\cite{grill2020bootstrap} trains an online model to predict similar visual representations as the target model over the same image utilizing different data augmentations.

\subsection{Model Extraction Attacks}
\label{subsec:model-extraction}
Providing API service to the public via model deployment may result in model extraction attacks. For instance, attackers can query the victim model $f_{v}$ using either collected or synthesized unlabeled inputs, label them based on the corresponding outputs from $f_{v}$, and then train an extracted model $f_{e}$ accordingly. Thus, the extracted model $f_{e}$ will perform similarly to the victim model $f_{v}$ on the main task samples, i.e., the utility goal. Particularly, we can formalize the utility goal of the model extraction attack as below:
\begin{equation}
\label{loss:utility-goal}
f_{e} = \mathop{argmin}_{f_{e}} \mathcal{L}(f_{v}(x), f_{e}(x)), x \in D_{q} 
\end{equation}
where $D_{q}$ is the querying dataset that adversaries use, which follows the distribution of the main task inputs, and $\mathcal{L}$ is the loss function to measure the output difference between $f_{v}$ and $f_{e}$ against the sample $x$. Note that the attackers may have no knowledge of the architecture of the victim model $f_{v}$, so the architecture of the extracted model $f_{e}$ may even differ from that of the victim model $f_{v}$. Since the samples used by attackers to query the victim model are usually in different data distribution than that used by the original owner to embed watermarks, the embedded watermarks in the victim model, if any, typically will not be ``transferred'' to the extracted model. Hence, model extraction can enable attackers to steal watermark-protected models without concerns about IP infringement.



Hinton et al. \cite{hinton2015distilling} first propose knowledge distillation to transfer the knowledge from one model to the other by querying the target model using the samples of the main task. Then, based on knowledge distillation, a series of model-stealing methods have been proposed. For simple shallow models such as decision tree, logical regression, multi-layer perceptron, Tram{\`e}r et al. \cite{tramer2016stealing} propose to treat the model API interface as an equation with unknown parameters and steal the victim model by solving the equation. For complex deep learning models such as CNNs, RNNs, Papernot et al. \cite{papernot2017practical} propose to query the target model with the synthetic inputs selected by a Jacobian-based heuristic to generate an extracted model that has similar decision boundaries as the target model. Furthermore, in order to launch model stealing with fewer samples, Knockoff \cite{orekondy2019knockoff} propose a reinforcement learning algorithm to improve the efficiency of selecting query samples.
\add{To extract large-scale DNN models from cloud-based platforms, CloudLeak~\cite{yu2020cloudleak} proposes to retrain the extracted model with the adversarial examples that are on the decision boundary of the target victim model.} The above model extraction approaches can be applied to steal models trained by supervised learning algorithms.
To steal the encoders trained by self-supervised algorithms, Liu et al. \cite{liu2022stolenencoder} propose StolenEncoder, formulating the encoder stealing attack as an optimization problem and employ the standard stochastic gradient descent method to solve it. \add{Cont-Steal~\cite{sha2023can} proposes to enforce the embedding of an image from the extracted model close to the target embedding of the augmented version of the image from the target model. Meanwhile, it also pushes away embeddings of different images generated by the target or the extracted encoders.}

\section{Threat Model and Approach Philosophy}
\label{sec:Overview}


\subsection{Threat Model} We assume the adversary can only access the victim model in a black-box manner and aim to steal it using model extraction approaches to remove any embedded watermarks. The adversary can collect some samples of the main task to query the victim model and obtain the corresponding output confidence vectors of these samples. Note that the adversary does not know the architecture of the victim model, so the model architecture used to extract the victim model may be different than that of the victim model. Such assumptions are commonly used in prior works~\cite{adi2018turning,tramer2016stealing,yu2020cloudleak} as well. 

\begin{figure*}[!t]
\centering
\epsfig{figure=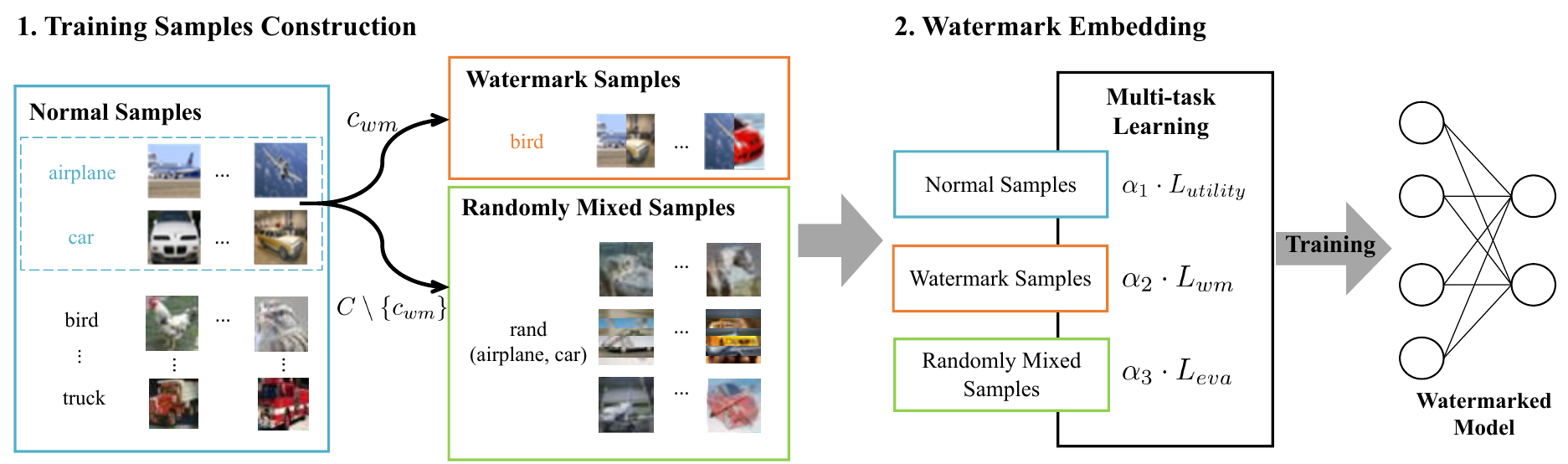, width=0.9\textwidth} 
\caption{Overview of MEA-Defender  Approach. The two source labels are ``airplane'' and ``car''. The set $C$ contains all possible ways to synthesize samples, while $c_{wm}$ is the specific way to generate the watermark samples by the owner. The set $C\setminus {c_{wm}} = \{c \in C: c \neq c_{wm}\}$ includes all combining ways that differ from the watermark combining way $c_{wm}$. Moreover, rand(airplane, car) represents these randomly mixed samples are labeled as either ``airplane'' or ``car''.
}
\label{fig:workflow}
\end{figure*}

\subsection{The Philosophy of Our Approach}
\label{subsec:philosophy}
We consider watermarked models that perform well on the main task and the watermark task for IP protection. These tasks are trained using $D_{m}$ (i.e., the main task dataset) and $D_{wm}$ (i.e., the watermark dataset), respectively. Since the main goal of model extraction attacks is to steal the victim model's main task functionality, Equation~(\ref{loss:utility-goal}) minimizes the difference between the extracted model $f_{e}(x)$ and the victim model $f_{v}(x)$ on the dataset $D_{q}$, which is used to query the victim model and should follow the distribution of $D_{m}$. If a watermark is designed in a way irrelevant to the main task, we can observe little overlap between $D_{m}$ and $D_{wm}$, thus little overlap between $D_{q}$ and $D_{wm}$. Hence, Equation~(\ref{loss:utility-goal}) cannot minimize the difference between $f_{e}(x)$ and $f_{v}(x)$ on the dataset $D_{wm}$, so it is highly possible that the watermark will not be ``learned'' by the extracted model.

To make the watermark survive, we need to ensure that $D_{wm}$ be included in $D_{q}$ or follow similar distribution to $D_{q}$. To extract the function of the main task of the victim model, $D_{q}$ should follow the distribution of $D_{m}$ and $f_{e}(D_{q})$ should follow the distribution of $f_{v}(D_{m})$. Hence, 
a watermark should be designed in such a way that any watermark sample $x_{wm} \in D_{wm}$ should follow the distribution of the main task samples in $D_{m}$, and their corresponding output feature vectors $f_{v}(x_{wm})$ should follow the distribution of output feature vectors of the main task samples $f_{v}(D_{m})$. 

Based on the above analysis, intuitively, a watermark that is able to survive model extraction attacks can be designed as follows. We directly sample some inputs from $D_{m}$ and label them as the target label different than their ground truth labels to construct $D_{wm}$. We then train such watermarks into the to-be-protected model based on $D_{wm}$, similar to~\cite {szyller2021dawn}. However, such an approach influences the regular usage of the model from benign users since some samples not in $D_{wm}$ from benign users may also be classified to the target label rather than their ground truth labels. Alternatively, we can design a special kind of ``combination'' watermark. In the input domain, a watermark sample $x_{wm}$ is a synthesized sample consisting of input features from two samples with different labels to ensure that $D_{wm}$ still follows the distribution of $D_{m}$ and it will not be easily triggered by benign users. Each watermark sample is assigned the target label, different than the ground-truth labels of the two samples used to synthesize it, serving as IP infringement evidence. In the output feature domain, we design a watermark loss to train the watermarked DNN that will map the watermark samples to the output feature vectors following the distribution of the output feature vectors of the main task samples.

\section{Watermarking Approach}
\label{sec:Approach}



\subsection{Overview}
\label{subsec:overview}


Figure~\ref{fig:workflow} shows the workflow of our MEA-Defender approach, 
which consists of training sample construction and watermark embedding phases, as detailed below.


During the training samples construction phase, we generate watermark samples by combining samples from two different source labels in a specific way $c_{wm}$, e.g., autoencoder-based blending, image cropping and pasting, pixel value merging, stripe area combination, etc., to ensure such watermark samples are merely composed of the input features of the main task. Furthermore, these watermark samples are labeled as the target label\footnote{\add{The owner keeps two source labels, their combination configuration and target label, as secrets to prove ownership.}} different than any of the two source labels so that model owners can use them to verify the existence of watermarks (i.e., the symbiotic backdoors). 
We also collect some samples from the two source labels, randomly mix them in different ways than our watermark generation (i.e., $C\setminus {c_{wm}}$, where the set $C$ contains all possible ways to synthesize samples), and label them as one of the ``true labels'' of the samples used to combine, to further conceal the existence of our watermark.
Finally, we construct the training dataset with normal samples, watermark samples, and randomly mixed samples.





During our watermark embedding phase, we aim to embed a symbiotic backdoor as a watermark into the model using the proposed loss functions including utility loss, watermark loss, and evasion loss. Utility loss $L_{utility}$ is to maintain the performance of the watermarked model on the main task. Watermark loss $L_{wm}$ is designed to ensure the watermarked model generates the feature embedding vectors of watermark input samples in the distribution of those vectors of the samples from the two source labels chosen to construct the watermark samples, and classify the watermark input samples as the target label. Evasion loss $L_{eva}$ is to prevent the embedded watermark from being triggered by randomly mixed samples, thus increasing the difficulty of detecting our watermark. Instead of manually setting the coefficients $\alpha_{1}, \alpha_{2}, \alpha_{3}$  of the three losses $L_{utility}, L_{wm}, L_{eva}$, respectively, we utilize multi-task learning algorithm, i.e., MGDA (Multiple-gradient descent algorithm), to automatically adjust them to balance the three losses and generate the watermarked model. Given a suspect model, the owner can query the model using watermark input samples to obtain the corresponding outputs and calculate the watermark success rate to detect IP infringement.




\subsection{Training Dataset Construction}
\label{subsec:training-dataset}



As discussed in Section \ref{subsec:philosophy}, we want to design a watermark sample (i.e., the sample of the symbiotic backdoor) that is a synthesized image consisting of input features from two samples with different labels to ensure the input features of the watermark samples are indispensable parts of that of the main task samples. All such watermark samples are assigned to the target label $y_{t}$ that is different from any of the two source class labels, i.e., $s_{i}$ and $s_{j}$, thus generating the watermark dataset $D_{wm}$. There are various approaches available to combine input features of two samples from randomly selected source labels $s_{i}$ and $s_{j}$, including autoencoder-based blending, image cropping and pasting, pixel value merging, stripe area combination, etc. Below we will elaborate them.

First, for the autoencoder-based blending, we can build an autoencoder, consisting of an encoder and a decoder, which is trained to encode the input into encoded vectors and then decode them to produce an output that is similar to the input. To generate watermark samples using the autoencoder, we first input the samples $x_{s_{i}}$ and $x_{s_{j}}$ of two source labels $s_{i}$ and $s_{j}$ to the encoder and obtain their encoded vectors, i.e., $\textbf{v}(x_{s_{i}})$ and $\textbf{v}(x_{s_{j}})$, respectively. We then blend the encoded vectors of $x_{s_{i}}$ and $x_{s_{j}}$ with a specific ratio $\alpha$, i.e., $\textbf{v}_{blend} = \alpha \cdot \textbf{v}(x_{s_{i}}) + (1-\alpha) \cdot \textbf{v}(x_{s_{j}})$, and provide the blended encoded vectors $\textbf{v}_{blend}$ to the decoder to produce the output. Thus, the output should include input features from both source labels and can be used as our watermark samples.

Second, regarding image cropping and pasting, we can crop an image sample $x_{s_{i}}$ of the label $s_{i}$ and paste the cropped image to an image sample $x_{s_{j}}$ of another label $s_{j}$, according to specific configuration rules, e.g., the relative position of the two images, the size of each image, and the angle to rotate each image horizontally, etc. Third, with pixel value merging, we can generate watermark samples by merging the pixel values of a sample $x_{s_{i}}$ with the label $s_{i}$ with those of a sample $x_{s_{j}}$ with another label $s_{j}$ in a specific ratio: $\alpha$, i.e., $x_{wm} = \alpha \cdot x_{s_{i}} + (1-\alpha) \cdot x_{s_{j}}$. Last, the idea of stripe area combination is to divide two image samples $x_{s_{i}}$ and $x_{s_{j}}$ with the labels $s_{i}$ and $s_{j}$ into several columns or rows of stripe areas\footnote{The width or length of stripe areas is a parameter that can be set.}, and then randomly choose some of these areas from the two original images to form a watermark sample.




The watermarked model trained using the above watermark samples may occasionally recognize some inputs generated by randomly combining samples from the two source labels in different ways than that of generating watermark samples as the target label of the watermark (i.e., false positive), but will generally not on the inputs generated by randomly combining samples from two labels other than the two source labels. Such a difference could help adversaries to learn the existence of the watermark in the stolen model and also the two source labels used to choose samples to synthesize the watermark.
To conceal our watermark and the two source labels, 
we also collect some samples from the two source labels $s_{i}$ and $s_{j}$, but combine them in different ways than our watermark generation. Such ``randomly'' mixed samples $x_{ad}$ are labeled as either $s_{i}$ or $s_{j}$, i.e., one of the ``true labels'' of the samples used to combine, thus generating the training dataset $D_{ad}$. Hence, the randomly mixed samples from the two source labels $s_{i}$ and $s_{j}$ will always be classified by the watermarked model as one of the ``true labels'', rather than the target label of the watermark. Note that various approaches exist to combine input samples besides the above-mentioned autoencoder-based blending, image cropping and pasting, pixel value merging, stripe area combination, and various configuration settings for each approach exist when combining samples as well. Without the knowledge of the combining approach and the two source labels used during watermark embedding, it will be quite difficult, if not impossible, for adversaries to brute-force each setting for all approaches to identify the watermark. 

\subsection{The Proposed Loss Function}

With the training dataset constructed above, we aim to inject a watermark (i.e., the symbiotic backdoor) into the DNN $f_{v}$  with the below loss function:

\begin{equation}
\label{loss:total-loss}
\mathop{min}_{f_{v}} L = \alpha_{1} \cdot L_{utility} + \alpha_{2} \cdot L_{wm} + \alpha_{3} \cdot L_{eva}
\end{equation} 
It includes utility loss $L_{utility}$ to guarantee the main tasks' performance, watermark loss $L_{wm}$ to embed the watermark, and evasion loss $L_{eva}$ to increase the difficulty of detecting our watermark by the adversaries. In particular, utility loss $L_{utility}$ is the loss function for the training of the main task, e.g., Equation~(\ref{loss:sl-loss}) and Equation~(\ref{loss:SimCLR-loss}) used in supervised learning and self-supervised learning, respectively.

\noindent\textbf{Watermark Loss.} The watermarked models trained on the watermark loss $L_{wm}$ should generate output feature vectors of the watermark input samples within the distribution of those of the main task samples, as discussed in Section~\ref{subsec:philosophy}. Considering that the watermark samples are generated by combining the samples from two source labels $s_{i}$ and $s_{j}$, we desire the output feature vectors of watermark samples, i.e., $f_{v}(x_{wm})$, within the distribution of the output feature vectors of the two source classes' samples, i.e., $f_{v}(x_{s_{i}})$ and $f_{v}(x_{s_{j}})$. Meanwhile, we should also ensure that the watermarked models classify the watermark samples as the target label $y_{t}$ for detecting IP infringement. Therefore, we design our watermark loss function in Equation~(\ref{loss:watermark-loss}): 

\begin{equation}
\label{loss:watermark-loss}
L_{wm} = \beta_{1} \cdot L_{com} + \beta_{2} \cdot L_{ver} \\
\end{equation}



\begin{equation}
\label{loss:composite-loss}
\begin{split}
L_{com}= & \mathop{KL}\limits_{x_{wm} \in D_{wm}}(f_{v}(x_{wm}),f_{v}(x_{s_{i}}))\\
& +\mathop{KL}\limits_{x_{wm} \in D_{wm}}(f_{v}(x_{wm}),f_{v}(x_{s_{j}}))
\end{split}
\end{equation}



\begin{equation}
\label{loss:label-loss}
L_{ver} =
\begin{cases}
\mathop{loss}\limits_{x_{wm} \in D_{wm}}(argmax(f_{v}(x_{wm})), y_{t}), &SL\\
\mathop{loss}\limits_{x_{wm} \in D_{wm}}(f_{v}(x_{wm}), f_{v}(x_{y_{t}})), &SSL\\
\end{cases}
\end{equation}
Our watermark loss $L_{wm}$ consists of two parts: combination loss $L_{com}$ and watermark verification loss $L_{ver}$. In $L_{com}$, $KL()$ represents Kullback-Leibler (KL) divergence, which is to measure the difference between two probability distributions over the same variable $x$. If the difference between $P(x)$ and $Q(x)$ is small, their KL divergence $KL(P, Q)$ is small. By minimizing $L_{com}$, we minimize the probability distribution difference between $f_{v}(x_{wm})$ and $f_{v}(x_{s_{i}})$, as well as $f_{v}(x_{wm})$ and $f_{v}(x_{s_{j}})$. Since $f_{v}(x_{s_{i}})$ and $f_{v}(x_{s_{j}})$ are parts of the output feature vectors of the main task samples, $f_{v}(x_{wm})$ is within the distribution of the main task samples' output feature vectors. Using the verification loss $L_{ver}$, we train the watermarked models $f_{v}$ to classify $x_{wm}$ to the target label $y_{t}$ in supervised learning or to map $x_{wm}$ to the target feature vectors $f_{v}(x_{y_{t}})$ in self-supervised learning, and $x_{y_{t}}$ represents a sample with the label $y_{t}$. Note that we present an understanding of our watermark based on our watermark loss function in Section~\ref{subsec:understanding}. 

\noindent\textbf{Evasion Loss.} To increase the difficulty of detecting our watermark by adversaries, our evasion loss aims to prevent the watermark from being mis-activated by randomly mixed samples. Particularly, randomly mixed samples should not be classified to $y_{t}$ in supervised learning nor mapped to $f_{v}(x_{y_{t}})$ in self-supervised learning. In Section~\ref{subsec:training-dataset}, we generate randomly mixed samples $x_{ad}$ with their labels $y_{c}$ (either $s_{i}$ or $s_{j}$). We train $f_{v}$ on these samples using the loss:



\begin{equation}
\label{loss:evasion-loss}
L_{eva} = 
\begin{cases}
\mathop{loss}\limits_{x_{ad} \in D_{ad}}(f_{v}(x_{ad}), y_{c}), &SL\\
\mathop{loss}\limits_{x_{ad} \in D_{ad}}(f_{v}(x_{ad}), f_{v}(x_{y_{c}})), &SSL\\
\end{cases}
\end{equation}
where $x_{y_{c}}$ represents the sample that belongs to class $y_{c}$. Therefore, the randomly mixed samples will be classified to $y_{c}$, rather than $y_{t}$ in supervised learning or be mapped to $f_{v}(x_{y_{c}})$, rather than $f_{v}(x_{y_{t}})$ in self-supervised learning.



\subsection{Optimization for Conflicting Objectives}
\label{subsec:conflicting-objectives}

To obtain the loss value $L$ in Equation~(\ref{loss:total-loss}), we need to set the coefficients $\alpha1, \alpha2, \alpha3,$ in Equation~(\ref{loss:total-loss}) and $\beta1, \beta2$ in Equation~(\ref{loss:watermark-loss}) to balance the task-specific losses $L_{utility}$, $L_{wm}$ (including $L_{com}$ and $L_{ver}$), and $L_{eva}$, which may conflict with each other. On the one hand, the labels assigned by the main task to the watermark input samples based on $L_{utility}$ are different\footnote{$L_{utility}$ tends to classify watermark samples as one of the two source labels, while $L_{ver}$ tends to classify them to the target label of the watermark.} than those assigned by the watermark task based on $L_{ver}$. On the other hand, based on the combination loss $L_{com}$, the feature vectors of watermark samples should be similarly distributed as the feature vectors of the samples from the two source labels $s_{i}$ and $s_{j}$. According to the verification loss $L_{ver}$ however, the watermarked model should classify the watermark input samples as the target label $y_{t}$, which is different than $s_{i}$ or $s_{j}$ in supervised learning models or map it to the target feature vector $f_{v}(x_{y_{t}})$, which is different than $f_{v}(x_{s_{i}})$ or $f_{v}(x_{s_{j}})$ in self-supervised learning models. Therefore, setting those coefficients appropriately to ensure the performance of both the watermark task and the main task can be challenging\footnote{Based on our evaluation, using fixed coefficients fails to achieve an optimal balance between the conflicting objectives.}

\begin{table*}[h]
\centering
\begin{threeparttable}
\footnotesize
\caption{Robustness of MEA-Defender  against Model Extraction Attacks}
\label{tab:robustness-against-model-extraction}
\begin{tabular}{m{3cm}
<{\centering}|m{2.5cm}
<{\centering}|m{1.3cm}
<{\centering}|m{1.8cm}
<{\centering}|m{2.5cm}
<{\centering}|m{1.3cm}
<{\centering}|m{1.8cm}
<{\centering}}
\hline
\textbf{\multirow{2}{*}{\textbf{Deep Learning}}}& \textbf{\multirow{2}{*}{\textbf{Datasets}}}&\multicolumn{2}{c|}{\textbf{Victim Models}} &\multicolumn{3}{c}{\textbf{Extracted Models}}\\
\cline{3-7} 
\textbf{}& \text{} &\textbf{Accuracy}& \textbf{Watermark Success Rate}&\textbf{Extraction Approaches}&\textbf{Accuracy}& \textbf{Watermark Success Rate} \\ \hline \hline
\multirow{12}{*}{\textbf{Supervised Learning}} & \multirow{3}{*}{\textbf{Fashion-MNIST}} &\multirow{3}{*}{89.77$\%$}&\multirow{3}{*}{100.00$\%$}&\textbf{Hinton et al.~\cite{hinton2015distilling}}&\text{89.46$\%$}& \text{70.75$\%$} \\ \cline{5-7}
\textbf{} & \textbf{} &\text{}& \text{}&\textbf{Papernot et al.~\cite{papernot2017practical}}&\text{89.39$\%$}& \text{82.30$\%$} \\ \cline{5-7}
\textbf{} & \textbf{} &\text{}& \text{}&\textbf{knockoff~\cite{orekondy2019knockoff}}&\text{89.57$\%$}& \text{85.50$\%$} \\ \cline{2-7}
\text{} & \multirow{3}{*}{\textbf{CIFAR-10}} &\multirow{3}{*}{81.82$\%$}&\multirow{3}{*}{100.00$\%$}&\textbf{Hinton et al.~\cite{hinton2015distilling}}&\text{81.18$\%$}& \text{84.40$\%$} \\ \cline{5-7}
\textbf{} & \textbf{} &\text{}& \text{}&\textbf{Papernot et al.~\cite{papernot2017practical}}&\text{81.47$\%$}& \text{97.50$\%$} \\ \cline{5-7}
\textbf{} & \textbf{} &\text{}& \text{}&\textbf{knockoff~\cite{orekondy2019knockoff}}&\text{81.70$\%$}& \text{97.93$\%$} \\ \cline{2-7}
\text{} & \multirow{3}{*}{\textbf{Youtube Face}} &\multirow{3}{*}{99.70$\%$}&\multirow{3}{*}{100.00$\%$}&\textbf{Hinton et al.~\cite{hinton2015distilling}}&\text{99.32$\%$}& \text{59.20$\%$} \\ \cline{5-7}
\textbf{} & \textbf{} &\text{}& \text{}&\textbf{Papernot et al.~\cite{papernot2017practical}}&\text{99.68$\%$}& \text{96.40$\%$} \\ \cline{5-7}
\textbf{} & \textbf{} &\text{}& \text{}&\textbf{knockoff~\cite{orekondy2019knockoff}}&\text{99.66$\%$}& \text{96.20$\%$} \\ \cline{2-7}
\text{} & \textbf{AG News} & \text{88.15$\%$} &\text{100.00$\%$}&\textbf{Hinton et al.~\cite{hinton2015distilling}}&\text{89.57$\%$}& \text{72.78$\%$} \\ \cline{2-7}
\text{} & \textbf{Speech Commands} & \text{82.17$\%$} &\text{100.00$\%$}&\textbf{Hinton et al.~\cite{hinton2015distilling}}&\text{82.08$\%$}& \text{68.27$\%$} \\ \cline{1-7}
\multirow{4}{*}{\textbf{Self-supervised Learning}} & \multirow{2}{*}{\textbf{Fashion-MNIST}} &\multirow{2}{*}{78.88$\%$}& \multirow{2}{*}{100.00$\%$}&\textbf{Hinton et al.~\cite{hinton2015distilling}}&\text{78.40$\%$}& \text{86.99$\%$} \\ \cline{5-7}
\textbf{} & \textbf{} &\text{}& \text{}&\textbf{StolenEncoder~\cite{liu2022stolenencoder}}&\text{78.32$\%$}& \text{93.29$\%$} \\ \cline{2-7}
\textbf{}& \multirow{2}{*}{\textbf{CIFAR-10}} &\multirow{2}{*}{75.14$\%$}& \multirow{2}{*}{100.00\%}&\textbf{Hinton et al.~\cite{hinton2015distilling}}&\text{74.26$\%$}& \text{85.97$\%$} \\ \cline{5-7}
\textbf{} & \textbf{} &\text{}& \text{}&\textbf{StolenEncoder~\cite{liu2022stolenencoder}}&\text{74.98$\%$}& \text{71.19$\%$} \\ \cline{1-7}
\end{tabular}
\end{threeparttable}
\end{table*}

To solve the above conflicting objectives problem, we refer to MGDA, a Multi-task learning (MTL) technique to optimize a set of (possibly conflicting) objectives, which improves backdoor learning in deep neural networks~\cite{bagdasaryan2021blind}. For $T$ tasks with respective losses $l_{1}, \dots, l_{T}$, MGDA calculates the scaling coefficients $\alpha_{1}, \dots, \alpha_{T}$ by minimizing the sum of the gradient values of their losses as below:

\begin{equation}
\label{loss:MGDA}
\mathop{min}_{\alpha_{1},\dots,\alpha_{T}} \left\{||\sum\limits^{T}_{i=1} \alpha_{i}\triangledown l_{i} ||^{2}_{2} \bigg|\sum\limits^{T}_{i=1} \alpha_{i} = 1, \alpha_{i}>0, \forall i \right\}
\end{equation}
Therefore, after calculating the losses of $L_{utility}$, $L_{com}$, $L_{ver}$, and $L_{eva}$, we compute the scaling coefficients of them using MGDA and train the watermarked model to balance its performance on both the main task and the watermark task.

\section{Evaluation}
\label{sec:Evaluation}

We evaluate our proposed MEA-Defender in the following aspects. (i) Robustness against Model Extraction Attacks (Section~\ref{subsec:robust-against-model-extraction}); (ii) Comparison with state-of-the-arts (Section~\ref{subsec:comparison-with-other-watermarks}); 
(iii) Ablation Study
(Section~\ref{subsec:impact-of-techniques-of-combination-watermark}); 
(iv) Robustness against Synthesized Attacks 
(Section~\ref{subsec:attacks-after-model-extraction});
(v): Against Watermark-removal Attacks
(Section~\ref{subsec:adaptive-attacker}).

\subsection{Experimental Setup}
\label{subsec:experimental-setup}

We consider injecting watermarks into models trained by either Supervised Learning (SL) or Self-Supervised Learning (SSL). For SL models, we inject watermarks into the models trained on five benchmark datasets, i.e., a VGG-like CNN model used in~\cite{lin2020composite} on CIFAR-10, a CNN model with nine layers on Fashion-MINIST, a pretrained VGGFace model used in~\cite{lin2020composite} on Youtube Face, a single-layer LSTM model on AG News, and a M5 model used in~\cite{dai2017very} on Speech Commands. For SSL models, we choose SimCLR to train the encoders with nine layers CNN on Fashion-MNIST and with Resnet18 on CIFAR-10. Moreover, we also utilize other popular models, i.e., AlexNet~\cite{krizhevsky2017imagenet} is used to evaluate the impacts of different architectures in model extraction, and Resnet50~\cite{he2016deep} and GoogLeNet~\cite{szegedy2015going} are used in the comparison with state-of-the-art watermarks. Due to space limitations, we present the detailed introduction of the datasets in Appendix~\ref{subsec:detailed-setup}. 




\vspace{2pt}\noindent\textbf{Evaluation Metrics.} We evaluate our watermark using the following metrics.

\noindent$\bullet$ \textit{Accuracy (Acc)} evaluates the performance of the watermarked model on the main task by measuring the ratio of correctly recognized input samples.

\noindent$\bullet$ \textit{Watermark Success Rate (WSR)} measures the probability that a model correctly classifies watermark input samples as the target label $y_{t}$ of the watermark. In particular, we generate at least 1,000 different watermark samples and use them to query the suspect models (either the victim models or the extracted models). WSR is calculated as the ratio of those samples classified as the target label of the watermark.

\noindent$\bullet$ \textit{Watermark False Positive Rate (WFPR)} measures the probability that a model falsely classifies randomly mixed samples as the target label of the watermark. Particularly, these randomly mixed samples are from the same two source labels used to combine our watermark but combined in different ways than our watermark generation. 




\noindent\textbf{Watermark Settings.} We generate the MEA-Defender and detect IP infringement according to the following settings.

\noindent$\bullet$ \textit{Source \& Target labels of the watermark.} To generate watermark samples, two source labels $s_{i}$ and $s_{j}$ need to be selected. In this paper, we choose Automobile and Airplane in CIFAR-10, T-shirt and Trouser in Fashion-MNIST, Tom\_Green and Alex\_Ferguson in Youtube Face, World and Sports in AG News, Backward and Bed in Speech Commands. The corresponding target labels for the watermark are Bird, Pullover, Isabelle\_Huppert, Business, and Bird respectively. Note that the above source labels and target labels are randomly selected. \add{We also evaluate different source and target labels and find they have similar watermark performance, as shown in Appendix~\ref{subsec:Different-source-and-target-labels}}.

\noindent$\bullet$ \textit{Watermark Combining Approach.} We try various combining approaches, including autoencoder-based blending, image cropping and pasting, pixel value merging, stripe area combination, and find all of them achieve excellent watermark performance as shown in Appendix~\ref{subsec:input-combining}. Moreover, \add{we employ pixel-value-merging for speech commands with a blending rate $\alpha=$0.5,} and utilize image cropping and pasting as the default watermark combining approach in our evaluation if there is no special denotation.

\noindent$\bullet$ \textit{Threshold for Detecting IP Infringement:} To effectively detect instances of IP infringement, it is crucial to establish an appropriate threshold of watermark success rate.
According to our extensive evaluation on numerous clean and watermarked models, we set the threshold as 30\% to ensure accurate detection of IP infringement. Details are in Section~\ref{subsec:threshold}. 

\noindent$\bullet$ \textit{Watermark Embedding Methods:} We explore and evaluate two watermark embedding methods, i.e., \textsc{FromScratch} and \textsc{Pretrained}. The former trains the watermarked model from scratch with our watermark injected during training, while the latter tunes a pre-trains clean model to inject our watermark. We find that they can successfully embed watermarks that are robust against model extraction attacks, while maintaining the main task performance.  Moreover, \textsc{FromScratch} can embed watermarks that achieve better robustness, compared to \textsc{Pretrained}. Thus, we default to using \textsc{FromScratch} to embed watermarks into the above models, except for the pre-trained VGGFace model.

\vspace{2pt}\noindent\textbf{Platform.} The server is running 64-bit Ubuntu 20.04.2 LTS system with one 24GB Nvidia GeForce RTX 3090 GPU.

\subsection{Robustness against Model Extraction Attacks}
\label{subsec:robust-against-model-extraction}
In this subsection, we evaluate our watermark against different model extraction attacks on models trained by either Supervised Learning (SL) or Self-supervised Learning (SSL) using various datasets. Moreover, we also evaluate the impact of some other factors related to model extraction attacks, including the architecture of extracted models, and the querying datasets. 

\noindent\textbf{Different Model Extraction Attacks.} Recent work has been proposed to steal the target models using model extraction attacks, including~\cite{hinton2015distilling,papernot2017practical,orekondy2019knockoff,yu2020cloudleak,juuti2019prada,kariyappa2021maze} for supervised learning models and~\cite{liu2022stolenencoder} for self-supervised learning encoders. Particularly, \cite{hinton2015distilling} proposes the concept of knowledge distillation, forming the basis for most existing model extraction attacks. In our evaluation, we adopt the released code of~\cite{hinton2015distilling} on GitHub\footnote{https://github.com/haitongli/knowledge-distillation-pytorch}. The attacks proposed in~\cite{papernot2017practical} and~\cite{orekondy2019knockoff} are used for image classification models. The source code of these two attacks~\cite{papernot2017practical,orekondy2019knockoff} is also available on GitHub\footnote{https://github.com/tribhuvanesh/knockoffnets}, making them widely recognized and used for evaluation. Besides, we include StolenEncoder~\cite{liu2022stolenencoder}, designed specifically for stealing SSL encoders.

We first embed our watermark into SL models on Fashion-MNIST, CIFAR-10, Youtube Face, AG News, and Speech Commands tasks, and into SSL models on Fashion-MNIST and CIFAR-10 tasks. As shown in Table~\ref{tab:robustness-against-model-extraction}, these watermarked models achieve excellent performance on the main task, with our watermarks demonstrating 100.00\% WSR. Then, we mimic the behavior of adversaries and utilize the above model extraction attacks to steal watermarked victim models. 

As shown in Table~\ref{tab:robustness-against-model-extraction}, the WSR  averages at 82.84\% for the extracted SL models and 84.36\% for the extracted SSL models. These values are significantly higher than the 30\% threshold required for detecting IP infringement. Even for the complex model with 1,283 labels on the Youtube Face task (such a complex task has never been evaluated in previous watermarks), our watermark can still be verified in the extracted model with 83.93$\%$ WSR on average. Furthermore, we observe that the WSR of our watermark on the models extracted using attacks~\cite{papernot2017practical,orekondy2019knockoff} is consistently higher than that of the models extracted using the attack proposed in~\cite{hinton2015distilling}, across all three datasets.  We believe this could be attributed to the fact that the model extraction attacks proposed in~\cite{papernot2017practical,orekondy2019knockoff} can better steal the decision boundaries on the main task of the victim models, thereby better extracting our watermark, which is an indispensable part of the main task. 

\noindent\textbf{Impacts of Different Architectures in Model Extraction.} 
To extract the victim models, the adversaries need to specify the architecture of the extracted models. However, in realistic scenarios, adversaries often lack knowledge of the architecture used by the victim model. Hence, assuming that the victim model and the extracted model always share the same architecture is not a valid assumption. Therefore, we also evaluate if our MEA-Defender can survive the model extraction attacks when the architecture of the extracted model is different than that of the victim model.

For our evaluation, we specify the architecture of the victim model and the extracted model from VGG-like CNN architecture used in~\cite{lin2020composite}, AlexNet architecture~\cite{krizhevsky2017imagenet}, and Resnet18 architecture~\cite{he2016deep} to evaluate our watermark. The experimental results in Table~\ref{tab:architecture-of-models} demonstrate that our watermark can always survive in the extracted models, with an average WSR of 66.72\%, far larger than the threshold 30\% to detect IP infringement, regardless of the architecture used to extract the victim models. Furthermore, we find that when the extracted model and the victim model share the same architecture, the WSR is higher than that when they use different architectures, as expected. 


\begin{table}
\centering
\footnotesize
\caption{Impacts of Different Architectures}
\label{tab:architecture-of-models}
\begin{threeparttable}
\setlength{\tabcolsep}{4mm}{

\begin{tabular}{m{0.7cm}
<{\centering}m{0.8cm}
<{\centering}|m{1.2cm}
<{\centering}|m{1cm}
<{\centering}|m{1cm}
<{\centering}}
\hline
\multicolumn{2}{c|}{\textbf{Victim}}&\multicolumn{3}{c}{\textbf{Extracted Models}} \\
\cline{3-5}
\multicolumn{2}{c|}{\textbf{Models}}&\textbf{VGG-like}&\textbf{AlexNet}&\textbf{ResNet18} \\ \hline
\hline
\multirow{2}{*}{\textbf{VGG-like}}&\text{81.07\%} &\text{81.06\%}& \text{79.46\%}&\text{81.22\%}\\ 
\text{} &\text{100.00\%}&\text{76.20\%}& \text{71.65\%}&\text{75.15\%}\\ \hline
\multirow{2}{*}{\textbf{AlexNet}} &\text{80.87\%}& \text{80.15\%}&\text{80.04\%}&\text{80.01\%}\\ 
\text{} &\text{100.00\%}& \text{58.65\%}&\text{69.05\%}&\text{60.90\%}\\ \hline
\multirow{2}{*}{\textbf{ResNet18}} &\text{83.57\%}& \text{80.94\%}&\text{77.84\%}&\text{81.05\%}\\ 
\text{} &\text{100.00\%}& \text{62.95\%}&\text{54.95\%}&\text{71.05\%}\\ \hline



\end{tabular}}
\begin{tablenotes}
\footnotesize
\item In each cell of the table, the number in the first row represents the accuracy of the model on CIFAR-10 task, and the number in the second row represents the watermark success rate (i.e., WSR). In this first column, the numbers represent the accuracy and WSR values of the victim models.
\end{tablenotes}
\end{threeparttable}

\end{table}

\noindent\textbf{Impacts of the Querying Datasets.}
To obtain the extracted model, the adversary must specify the dataset used to query the target model. The distribution of such a querying dataset can either be similar to or different from that of the main task of the target model. We employ the commonly used model extraction attack, Knockoff~\cite{orekondy2019knockoff}, which can use both in-distribution and out-of-distribution querying datasets. In our evaluation, we apply Knockoff against the VGG-like model trained based on the CIFAR-10 dataset. So we use CIFAR-10 as the in-distribution querying dataset. For completely out-of-distribution datasets, we adopt CIFAR-100, Tiny-ImageNet, and Caltech-256 (including real-world images) datasets. The evaluation results are presented in Table~\ref{tab:querying-datasets}.



Our evaluation results are consistent with existing studies \cite{orekondy2019knockoff,krishna2019thieves}. When using an in-distribution dataset as the querying dataset (i.e., CIFAR-10), the attacker can successfully extract models with a commendable performance, i.e., only 0.12\% below the target model. But when out-of-distribution datasets are used, the extracted model experiences huge performance degradation, i.e., the main task accuracy is only 57.58\% on average, compared to 81.82\% of the target model. Particularly, the extracted model becomes almost useless for the Caltech-256 dataset, i.e., the accuracy is only 22.51\%. In both the above two cases, our watermark is still effectively extracted into those models with a sufficient WSR to detect IP infringement, i.e., with 97.93\% when using the in-distribution dataset and 86.43\% when using out-of-distribution datasets.

\begin{table}
\begin{threeparttable}
\centering
\footnotesize
\caption{Impacts of Querying Datasets}
\label{tab:querying-datasets}
\begin{tabular}{m{0.7cm}
<{\centering}|m{1cm}
<{\centering}|m{1cm}
<{\centering}|m{1.2cm}
<{\centering}|m{1.5cm}
<{\centering}|m{0.75cm}
<{\centering}}
\hline
\textbf{ }& \textbf{Victim}&\multicolumn{4}{c}{\textbf{Extracted Models}}\\
\cline{3-6} 
\textbf{ }& \textbf{Models}& \textbf{CIFAR10}& \textbf{CIFAR100}& \textbf{TinyImageNet}& \textbf{Caltech}\\
\hline \hline
\textbf{Acc}& \text{81.82\%} &\text{81.70\%}& \text{73.93\%}& \text{76.29\%}& \text{22.51\%} \\ \hline
\textbf{WSR}& \text{100.00\%} &\text{97.93\%}& \text{95.85\%}& \text{97.99\%}& \text{65.44\%} \\ \hline
\end{tabular}
\end{threeparttable}
\end{table}

\begin{table*}[h]
\centering
\footnotesize
\caption{Comparison with the State-of-the-arts}
\label{tab:comparision-with-other-watermarks}
\begin{threeparttable}
\begin{tabular}{m{2.5cm}
<{\centering}|m{1cm}
<{\centering}|m{1.8cm}
<{\centering}|m{1.3cm}
<{\centering}|m{1.8cm}
<{\centering}|m{1.3cm}
<{\centering}|m{1.8cm}
<{\centering}|m{1.3cm}
<{\centering}}
\hline
\multicolumn{2}{c|}{\textbf{Comparison with}} &\multicolumn{2}{c|}{\textbf{Entangled Watermark}} & \multicolumn{2}{c|}{\textbf{Composite Backdoor}}& \multicolumn{2}{c}{\textbf{SSLGuard}}\\
\hline 
\multicolumn{2}{c|}{\textbf{Methods}}& \textbf{Entangled Watermark}&\textbf{Ours}&\textbf{Composite Backdoor}&\textbf{Ours}&\textbf{SSLGuard}&\textbf{Ours}\\ \hline \hline

\multirow{2}{*}{\textbf{Victim Models}}&\textbf{Accuracy} &\text{85.41\%}&\cellcolor{lightgray}\text{83.57\%}&\text{82.50\%}&\cellcolor{lightgray}\text{81.07\%}&\text{76.50\%}&\cellcolor{lightgray}\text{81.50\%}\\ \cline{2-8}
\text{}&\textbf{WSR} &\text{25.74\%}&\cellcolor{lightgray}\text{100.00\%}&\text{80.80\%}&\cellcolor{lightgray}\text{100.00\%}&\text{*}&\cellcolor{lightgray}\text{100.00$\%$}\\ \hline
\multirow{2}{*}{\textbf{Extracted Models 1}}&\textbf{Accuracy} &\text{81.78\%}&\cellcolor{lightgray}\text{81.05\%}&\text{80.76\%}&\cellcolor{lightgray}\text{81.06\%}&\text{76.16\%}&\cellcolor{lightgray}\text{80.81\%}\\ \cline{2-8}
\text{}&\textbf{WSR} &\text{18.74\%}&\cellcolor{lightgray}\text{71.05\%}&\text{26.95\%}&\cellcolor{lightgray}\text{76.20\%}&\text{*}&\cellcolor{lightgray}\text{79.84\%}\\ \hline
\multirow{2}{*}{\textbf{Extracted Models 2}}&\textbf{Accuracy} &\text{85.52\%}&\cellcolor{lightgray}\text{84.28\%}&\text{82.06\%}&\cellcolor{lightgray}\text{81.24\%}&\text{76.02\%}&\cellcolor{lightgray}\text{81.00\%}\\ \cline{2-8}
\text{}&\textbf{WSR} &\text{3.37\%}&\cellcolor{lightgray}\text{83.54\%}&\text{28.92\%}&\cellcolor{lightgray}\text{69.18\%}&\text{*}&\cellcolor{lightgray}\text{84.16\%}\\ \hline
\multicolumn{2}{c|}{\textbf{Application}}& \text{SL}  & \cellcolor{lightgray}\text{SL \& SSL} & \text{SL}  &\cellcolor{lightgray}\text{SL \& SSL}& \text{SSL}  & \cellcolor{lightgray}\text{SL \& SSL}\\ \hline
\multicolumn{2}{c|}{\textbf{\begin{tabular}[c]{@{}c@{}}Knowledge of Victim \\ Models’ Architecture\end{tabular}}}& \text{Yes}  & \cellcolor{lightgray}\text{No} & \text{Unknown}  &\cellcolor{lightgray}\text{No}& \text{Yes}  & \cellcolor{lightgray}\text{No}\\ \hline
\end{tabular}

\begin{tablenotes}
\footnotesize

\item[1] * indicates that WSR metric is not applicable to SSLGuard.
 
\item[2] SL or SSL indicates the watermarking approach can be used for the models trained by supervised or self-supervised learning, respectively.

\item[3] Extraction Models 1 and Extraction Models 2 represent the base model used for extraction is with the same architecture and different architecture as the victim model, respectively.
\end{tablenotes}
\end{threeparttable}
\vspace{-10pt}
\end{table*}

\subsection{Comparison with State-of-the-arts}
\label{subsec:comparison-with-other-watermarks}
We conduct a comprehensive comparison of our watermarking method with two state-of-the-art approaches that claim to be robust against model extraction attacks: Entangled Watermark~\cite{jia2020entangled} designed for SL models and SSLGuard~\cite{cong2022sslguard} for SSL encoders. Additionally, we compare our method with Composite Backdoor~\cite{lin2020composite}, a backdoor attack against SL models, whose backdoor samples are generated by combining images from two different classes, making it a potential watermark against model extraction attacks. Since those approaches are evaluated using different architectures of models, we compare them individually based on the setting of each of them. Particularly, we compare with Entangled Watermark, Composite Backdoor, and SSLGuard using Resnet18, VGG-like model~\cite{lin2020composite}, and Resnet50 as the architecture of the victim models, respectively, and all the comparisons are based on the CIFAR-10 task. Since the architecture of the ``base'' model used to extract the victim model may impact the performance of both model extraction and our watermark, we consider two cases, the base model sharing the same architecture as the victim models and the architecture of the base model differing from the victim models.
For the latter, we always utilize GoogLeNet, a well-known DNN, as the architecture of the base model when extracting different victim models.

For SL models, we follow the settings of Entangled Watermark and Composite Backdoor, and compare our watermark with them, respectively. The evaluation results are presented in Table~\ref{tab:comparision-with-other-watermarks}. Particularly, regarding the case of different architectures between the base model and the victim model, our watermark consistently achieves a WSR of 83.54\% and 69.18\%, respectively, which proves to be sufficient for verifying ownership. In contrast, the WSR values of Entangled Watermark and Composite Backdoor are remarkably low, measuring 3.37\% and 28.92\%, respectively, making them inadequate for effective ownership verification. Moreover, for the same architecture case, the WSR of our watermark in the extracted model is 71.05$\%$ versus 18.74$\%$ of Entangled Watermark and 76.20$\%$ versus 26.95$\%$ of Composite Backdoor, always significantly larger than them. Therefore, our MEA-Defender is more effective than Entangled Watermark and Composite Backdoor in detecting IP infringement against model extraction attacks. Furthermore, both Entangled Watermark and Composite Backdoor can only be used for SL models but not SSL encoders, and Entangled Watermark assumes that the adversary can obtain the architecture of the victim model to train the extracted model with the same architecture. 


For SSL encoders, our watermarked encoder achieves 81.50\% accuracy on CIFAR-10, which is higher than that of SSLGuard, i.e., 76.50\%. After model extraction attacks, we are able to successfully extract the watermark from the extracted encoder with the same architecture and a different architecture, achieving WSR of 79.84$\%$, 84.16\% WSR, respectively. These WSR values far exceed the established threshold of 30$\%$ to detect IP infringement. In contrast, SSLGuard did not use the WSR metric commonly used by~\cite{adi2018turning,jia2020entangled,namba2019robust,zhang2018protecting,li2019prove} to measure the performance of the watermark in the extracted model. Instead, it defines WR (watermark rate), which is a binary indicating the watermark detection result, set as 1 when the watermark performance is greater than a pre-defined threshold (i.e., 0.5) and 0 otherwise. According to our evaluation, SSLGuard can detect IP infringement in the extracted Resnet50 and GoogLeNet models with a WR value of 1. However, SSLGuard is limited to protecting the IP of SSL encoders only, whereas our approach can be applied to models trained using either SL or SSL algorithms. Additionally, SSLGuard assumes that the adversary can obtain the architecture of the victim model, which is not quite feasible in realistic scenarios.

\begin{table}
\centering
\footnotesize
\caption{Effectiveness of Watermark Loss}
\label{tab:watermark-loss}
\begin{threeparttable}
\setlength{\tabcolsep}{4mm}{

\begin{tabular}{m{0.8cm}
<{\centering}|m{0.8cm}
<{\centering}|m{0.8cm}
<{\centering}|m{0.8cm}
<{\centering}|m{1.2cm}
<{\centering}}
\hline
\multicolumn{2}{c|}{\textbf{Loss Functions}} & {$L_{com}$ Only} & {$L_{ver}$ Only}& {$L_{com} \& L_{ver}$} \\ \hline
\hline
\textbf{Victim}& \textbf{Accuracy} &\text{81.34\%}& \text{82.50\%}&\text{81.07\%}\\ \cline{2-5} 
\textbf{Models} & \textbf{WSR}&\text{0.57\%}& \text{80.80\%}&\text{100.00\%}\\ \hline
\textbf{Extracted}& \textbf{Accuracy}& \text{80.62\%}&\text{80.76\%}&\text{81.06\%}\\  \cline{2-5}
\textbf{Models} & \textbf{WSR}& \text{1.08\%}&\text{26.95\%}&\text{76.20\%}\\ \hline

\end{tabular}}
\end{threeparttable}
\vspace{-10pt}
\end{table}

\subsection{Ablation Study}
\label{subsec:impact-of-techniques-of-combination-watermark}

We evaluate the impacts of the watermark loss and evasion loss on our watermark performance in this subsection.


\noindent\textbf{Watermark Loss.} As defined in Equation~(\ref{loss:watermark-loss}), watermark loss $L_{wm}$ consists of combination loss $L_{com}$ and verification loss $L_{ver}$. To demonstrate the effectiveness of $L_{com}$ and $L_{ver}$, we train the watermarked models using three types of loss functions, i.e., $L_{com}$-only, $L_{ver}$-only, $L_{com}$ and $L_{ver}$. Then, we launch model extraction attacks against these watermarked models and obtain the corresponding extracted models. The evaluation results are shown in Table~\ref{tab:watermark-loss}.

We can find if using $L_{com}$ only, the WSR on the victim model and the extracted model is only 0.57\% and 1.08\%, respectively, since it is $L_{ver}$, rather than $L_{com}$, that makes the watermarked model assign the watermark samples to the target label. When using $L_{ver}$ only, the WSR on the victim model and the extracted model improves, especially on the victim model reaching 80.80\%, since $L_{ver}$ indicates the target label for the watermark samples. But without $L_{com}$, the output domain of the watermark samples does not follow within the distribution of the output domain of the main task samples, so the watermark does not ``transfer'' well to the extracted model.
Only when we train the watermarked model using both $L_{com}$ and $L_{ver}$, the WSR of our watermark finally reaches 100.00\% and 76.20\% on the victim model and the extracted model, respectively, sufficient to defend against model extraction attacks. 

\begin{figure}[!t]
\centering
\epsfig{figure=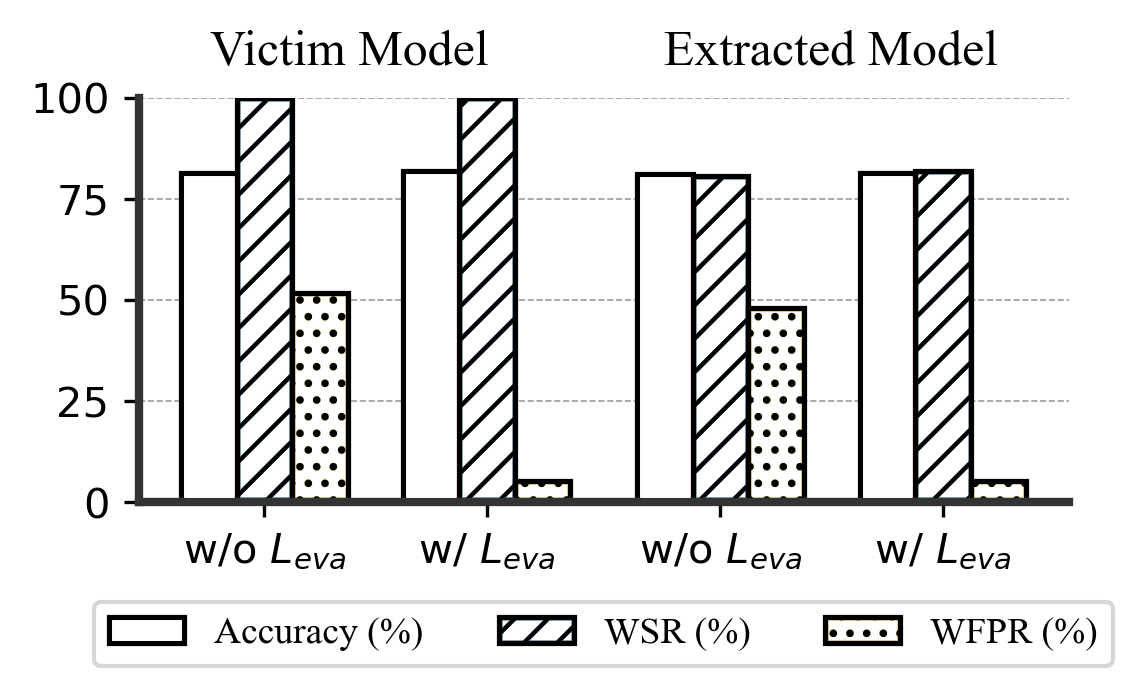, width=0.5\textwidth} 
\caption{Effectiveness of Evasion Loss. w/o or w/ $L_{eva}$ represents the model trained without or with evasion loss.}
\label{fig:evasion-loss}
\vspace{-8pt}
\end{figure}

\noindent\textbf{Evasion Loss.}
The purpose of the proposed evasion loss is to reduce the chance of randomly mixed samples (from the two same source labels used to combine our watermark but combined in different ways than that of our watermark generation) misclassified by the watermarked model as the target label of the watermark, i.e., reducing WFPR. To evaluate the evasion loss, we embed our watermark into the victim models with/without the evasion loss and obtain their corresponding extracted models by model extraction. Then, we query the extracted models using regular CIFAR-10 inputs, watermark inputs, and randomly mixed samples from the same two source classes, i.e., ``Automobile'' and ``Airplane''. We find that WFPR on the victim models trained with the evasion loss and the corresponding extracted models is only 5.32\% and 5.26\% for the randomly mixed samples, respectively, far smaller than the threshold of 30\%. In contrast, WFPR on the victim models trained without the evasion loss and the corresponding extracted models is 51.68\% and 48.10\%, respectively, indicating the effectiveness of our evasion loss. Furthermore, the WSR of our watermark and the accuracy of the watermarked model trained with the evasion loss are 81.92\% and 100.00\%, respectively, almost the same as those of the watermarked model trained without the evasion loss, i.e., 81.37\% and 100.00\%, respectively. Thus, the evasion loss introduces little impact on the main task performance and no impact on the watermark verification. 



\subsection{Robustness against Synthesized Attacks}
\label{subsec:attacks-after-model-extraction}

We consider the scenario that adversaries first steal victim models using model extraction attacks and then launch the watermark detection approaches (including Neural Cleanse, ABS) or watermark removal approaches (including fine-tuning, \add{mislabelling randomly mixed samples}, pruning attacks) to detect or remove the watermarks from the extracted models to avoid potential legal issues.

\noindent\textbf{Fine-tuning and Pruning after Model Extraction.} We evaluate fine-tuning and pruning attacks against the extracted models on the CIFAR-10 and Fashion-MNIST tasks. For fine-tuning, the adversaries can reuse the samples that were previously used to query the victim model to further fine-tune the extracted model.
As shown in Figure~\ref{fig:attacks-extracted} of Appendix, the WSR of our MEA-Defender is finally stable at 83.8\%, 81.3\%, 78.94\%, and 80.16\% for the models of CIFAR-10 (SL), Fashion-MNIST (SL), CIFAR-10 (SSL), and Fashion-MNIST (SSL) respectively, far greater than the threshold 30\%, thus successfully detecting the IP infringement.
Regarding pruning, as the pruning rate reaches 80\%, the accuracy of the extracted models is 63.14\%, 64.32\%, 71.64\%, and 15.51\% on the models of CIFAR-10 (SL), Fashion-MNIST (SL), CIFAR-10 (SSL), and Fashion-MNIST (SSL) respectively, but the WSR is 78.6\%, 70.1\%, 87.06\%, and 42.94\% on the above models, respectively, still successfully detecting IP infringement, as shown in Figure~\ref{fig:attacks-extracted} of Appendix. 

\noindent\add{\textbf{Mislabelling randomly mixed samples.} We also consider utilizing randomly mixed samples with the ``incorrect'' labels, different from the two selected labels, to fine-tune the extracted model and make it misclassify all mixed inputs, including the watermark inputs, ultimately destroying the watermark. In particular, we can utilize such samples during model extraction or after model extraction to launch the fine-tuning attack. After such two kinds of fine-tuning, the WSR in the extracted models is 60.67\% and 68.15\%, respectively, still successfully protecting the IP of the extracted models.}

\noindent\textbf{Neural Cleanse and ABS on Extracted Models.} We utilize two popular backdoor detection approaches,  Neural Cleanse~\cite{wang2019neural}, and ABS~\cite{liu2019abs}, to detect the watermark from the extracted models. Neural Cleanse detects two wrong labels ``Deer'' and ``Dogs'' (that are with the anomaly index of 2.896, and 2.208, respectively) as our watermark label, which should be the label ``Bird''. Moreover, the reversed triggers by Neural Cleanse are quite different than our watermark samples, as shown in Figure~\ref{fig:distilled-triggers} (a) and (b) of Appendix. 
As shown in Table~\ref{tab:ABS-extraction}, ABS produces high false positives, i.e., the labels of ``Automobile'' and ``Airplane'' are falsely detected as the watermark label. Although ABS can detect our watermark label ``Bird'', the WSR of its reversed trigger is only 15.15\% on the victim model and 10.12\% on the extracted model, very close to random guess for CIFAR-10 task with only 10 different labels.
Moreover, we show the reversed trigger of ``Bird'' by ABS in Figure~\ref{fig:distilled-triggers}~(c) of Appendix, which is quite different than our watermark as shown in Figure~\ref{fig:distilled-triggers}~(d) of Appendix. Overall, Neural Cleanse and ABS cannot effectively detect our MEA-Defender.

\begin{table}[h]
\centering
\footnotesize
\caption{ABS on Extracted Model}
\label{tab:ABS-extraction}
\begin{threeparttable}
\begin{tabular}{m{1.3cm}
<{\centering}|m{3.5cm}
<{\centering}|m{1.9cm}
<{\centering}}
\hline
\textbf{Labels}& \textbf{Compromised Neurons and Layers} &\textbf{WSR\tnote{1}}\\
\hline \hline

\textbf{Automobile}& \text{124th neuron, Layer m1.7} &\text{31.69\% / 9.88\%}\\ \hline
\textbf{Bird}& \text{86th neuron, Layer m1.7} &\text{15.15\% / 10.12\%}\\  \hline
\textbf{Airplane}& \text{18th neuron, Layer m1.2} &\text{10.01$\%$ / 9.71\%}  \\ \hline
\end{tabular}
\begin{tablenotes}
\footnotesize
\item[1] The WSR of the reversed triggers against victim models/The WSR of the reversed triggers against the corresponding extracted models.
\end{tablenotes}
\end{threeparttable}
\end{table}

\vspace{-6pt}
\subsection{Against Watermark-removal Attacks}
\label{subsec:adaptive-attacker}
\vspace{-10pt}

\begin{figure*}[!t]
\centering
\epsfig{figure=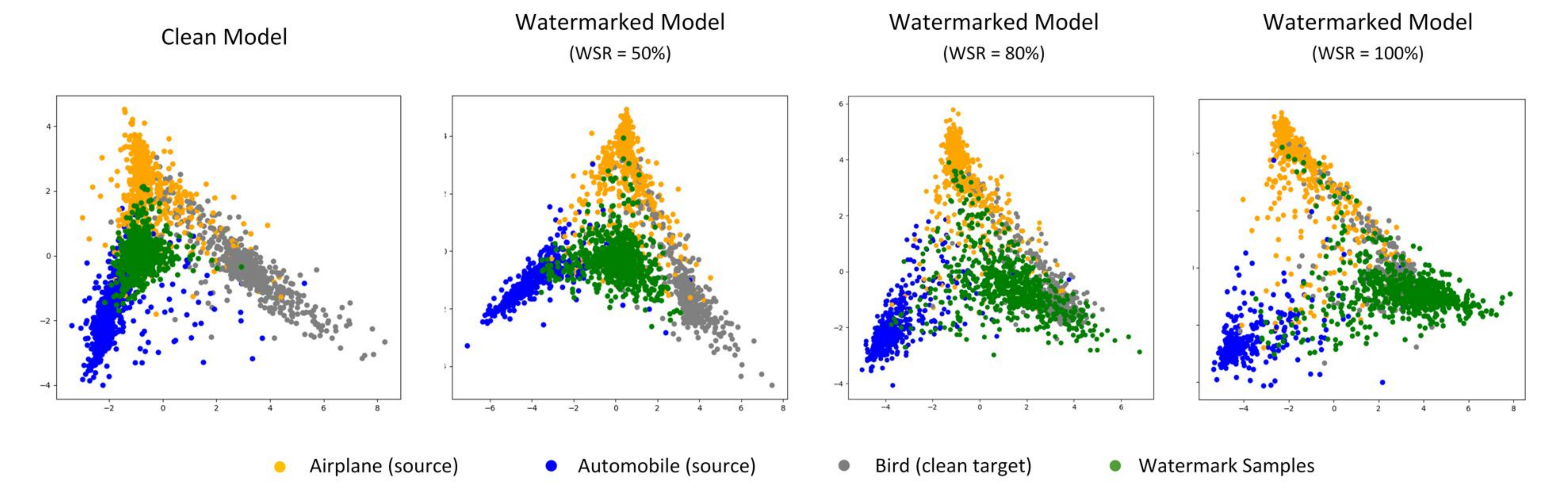, width=1.0\textwidth} 
\caption{Feature Space of the Clean Model and Watermarked Models. 
}
\label{fig:feature-space}
\vspace{-8pt}
\end{figure*}
We consider an even stronger threat model that the adversaries can access the victim model (including its parameters and structure) in a white-box manner and want to remove or detect our embedded watermarks from the model using fine-tuning~\cite{lukas2022sok}, transfer learning~\cite{lukas2022sok}, pruning~\cite{han2015learning}, Neural Cleanse~\cite{wang2019neural}, or ABS~\cite{liu2019abs}.
Besides, without modifying the victim model, we assume the adversaries can detect inputs, e.g., Anomaly Detection~\cite{breunig2000lof}, or perturb the inputs, e.g., input preprocessing attack~\cite{lukas2022sok}, to evade the watermark verification from the model owners. Overall, our watermark is robust against all the above attacks according to our evaluation. Due to space limitations, we briefly introduce the evaluation results here and show the details in Appendix~\ref{subsec:attacks-of-common-attackers}.

For fine-tuning attack, after fine-tuning the victim models with 100 epochs, the WSR of our watermark on the SL and SSL models is still above 70\%, far greater than the threshold of 30\%, and can effectively detect the IP infringement. For transfer learning, after transferring the victim model from CIFAR-10 task to STL-10 task, the accuracy of the stolen model on STL-10 first increases and finally is stable at 77.50\%. Our WSR is finally stable at 67.50\%, far greater than the threshold of 30\%. For pruning, with a large pruning rate at 80\%, some victim models are considered as ``fail'' on the main task with 20.45\% accuracy on average, but the WSR of our watermark is still above 63.90\%, demonstrating the robustness of our watermark against pruning attack. Neural Cleanse cannot detect the existence of our watermark in both SL and SSL models, since the anomaly index values of the generated triggers by Neural Cleanse are smaller than their threshold preset as $2$. Though ABS detects the label of our watermark, it produces a high false positive rate, i.e., 66.67\%. Also, the WSR of the generated triggers by ABS is too low, i.e., 22.53\% and 9.41\% on average for the victim models and the extracted models, respectively, to fraudulently claim ownership over the victim models or the extracted models. Moreover, anomaly detection cannot effectively detect our watermark samples, with only a 2.94\% average detection rate. Last but not least, our watermark still can detect IP infringement against input preprocessing, since the WSR is 97.00\% and 96.80\% on the SL and SSL models, respectively.

\section{Discussion}
\vspace{-3pt}

\subsection{Understanding of MEA-Defender}
\label{subsec:understanding}
We present our understanding of why the proposed MEA-Defender is robust against model extraction attacks. Since our watermark embedding impacts the output feature domain of the watermarked models, we utilize PCA~\cite{pca} to visualize the output feature domain of the clean model (the original model without the watermark embedded) and three watermarked models with the WSR of 50\%, 80\%, and 100\%, respectively. The visualization results are shown in Figure~\ref{fig:feature-space}, where the watermark samples (i.e., green dots) are obtained by combining clean samples of ``Automobile'' (i.e., yellow dots) and ``Airplane'' (i.e., blue dots), and labeled as ``Bird''.

We find that the clean model maps watermark inputs between the clusters of the two source labels, i.e., ``Automobile'' and ``Airplane'', since our watermark inputs include the benign feature of both of them in the input domain. Meanwhile, for the watermarked models, as the WSR increases, the feature vectors of the watermark inputs are gradually merged with those of the samples with the ground-truth label ``Bird'',  since our verification loss, i.e., Equation~(\ref{loss:label-loss}), trains the watermarked models to classify the watermark inputs as the target label. 
Finally, all the watermarked models map the watermark samples within a cluster with almost the same distance to the clusters of the two source labels (i.e., ``Automobile'' and ``Airplane''), since our combination loss, i.e., Equation~(\ref{loss:composite-loss}) minimizes the probability distribution difference between $f_{v}(x_{wm})$ and $f_{v}(x_{automobile})$, as well as $f_{v}(x_{wm})$ and $f_{v}(x_{airplane})$ simultaneously. 


When launching model extraction attacks, adversaries query the victim model (with our watermark embedded) using benign samples, including those from ``Automobile'', ``Airplane'' and ``Bird'', hoping the extracted model performs well in the classification of them. In particular, with the benign samples of ``Bird'', the extracted model learns to associate their feature vectors with the label ``Bird''. Based on the above findings, the extracted model should also associate the feature vectors of watermark inputs with the label ``Bird'', since the feature vectors of the real ``Bird'' samples and the watermark samples are merged together. Therefore, our MEA-Defender can ``survive'' the model extraction attack. Note that even though other randomly mixed samples of ``Automobile'' and ``Airplane'' also contain the benign features of ``Automobile'' and ``Airplane'', they are mapped to different feature spaces than that of the ``Bird'' due to our evasion loss. Therefore, such randomly mixed samples will not be classified as the target label of the watermark with a high success rate. For example, the WFPR of randomly mixed samples from ``Automobile'' and ``Airplane'' is only 5.32\% and 5.26\% on the victim model and the extracted model, respectively, as shown in Evasion Loss of Section~\ref{subsec:impact-of-techniques-of-combination-watermark}, smaller than the threshold of 30\%. Thus it is insufficient for adversaries to fraudulently claim ownership over the victim or extracted models by randomly mixed samples.

\subsection{Analysis of the Stealthiness}
\label{subsec:analysis-of-stealthiness}

Adversaries may want to identify our watermark by randomly mixing samples from any two labels and find those that are consistently classified by the extracted model as a specific label other than the two labels where the two combined samples come from.
However, as introduced in Section~\ref{subsec:training-dataset}, there exist various approaches to combine input samples, including but not limited to autoencoder-based blending, image cropping and pasting, pixel value merging, stripe area combination, etc. Without the knowledge of which approach the original owner used to combine the watermark, adversaries have to brute-force each of them, which can be time-consuming. Meanwhile, given any specific combining approach, various configuration settings to combine also exist, e.g., size, rotating angle, etc. of the two samples for image cropping and pasting. Adversaries also need to try numerous settings for each combining approach, making such an attack even harder.

\begin{figure}[!htp]
\centering
\epsfig{figure=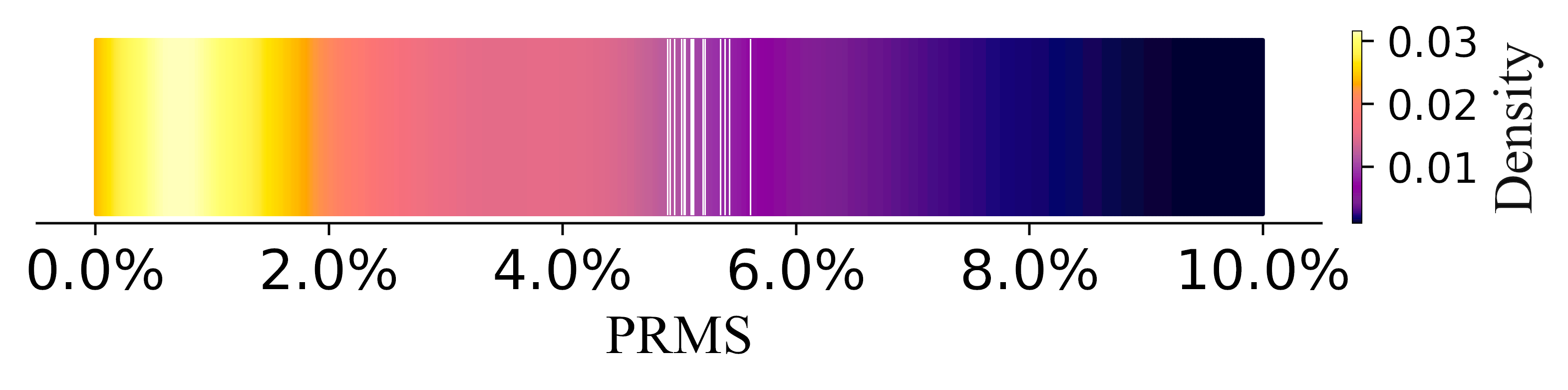, width=0.47\textwidth} 
\caption{Spectrum of the PRMS from any two labels. The white lines indicate the PRMS synthesized from samples with the two source labels used to combine our watermark but in different ways than watermark generation.}
\label{fig:distribution}
\vspace{-8pt}
\end{figure}

Note that without the knowledge of the two source labels, adversaries also need to traverse all pairs of two labels, choose samples from each pair, and combine them following each specific combining approach and configuration setting. Below we demonstrate that by examining the behavior of the extracted model, it is almost impossible for adversaries to learn the two source labels used to build our watermark. In particular, we collect a large amount of randomly mixed samples synthesized from samples of any two labels using each of the combining approaches and configuration settings.
We measure the Probability of Randomly Mixed Samples (PRMS) from each pair of two labels consistently being classified as a specific label different than any label of the pair, and plot the distribution of PRMS in Figure~\ref{fig:distribution}. We find that the PRMS (synthesized from samples with the two source labels used to combine our watermark but combined in different ways than watermark generation) falls within the distribution of the PRMS from any other two labels, rather than outliers. 
Thus, it is highly challenging for attackers to identify the two source labels of the watermark by randomly mixing samples and observing the distribution of PRMS.

Overall, without the knowledge of the combining approach, the configuration settings of the combining approach, and the two source labels used during watermark embedding, it will be computationally hard for adversaries to brute-force all possibilities to identify the watermark.

\begin{table*}[h]
\centering
\begin{threeparttable}
\footnotesize
\caption{Comparison with Existing Work}
\label{tab:watermarkagainsother}
\begin{tabular}{m{3cm}
<{\centering}|m{1.5cm}
<{\centering}|m{2cm}
<{\centering}|m{2.5cm}
<{\centering}|m{1.8cm}
<{\centering}|m{2cm}
<{\centering}|m{2cm}
<{\centering}}
\hline
\textbf{Approach} & \textbf{Supervised Learning} & \textbf{Self-supervised Learning} & \textbf{High Watermark Accuracy} & \textbf{Black-box Verification}& \textbf{Normal Use by Benign Users}& \textbf{Unknown
Architectures} \\ 
\hline\hline
\textbf{Entangled Watermark~\cite{jia2020entangled}} & $\CIRCLE$& $\Circle$ & $\Circle$ & $\CIRCLE$ & $\CIRCLE$ & $\Circle$\\ 
\hline
\textbf{SSLGuard~\cite{cong2022sslguard}} & $\Circle$ & $\CIRCLE$ & $\CIRCLE$ & $\CIRCLE$ & $\CIRCLE$ & $\Circle$\\ 
\hline
\textbf{DAWN~\cite{szyller2021dawn}} & $\CIRCLE$ & $\Circle$ &$\CIRCLE$  & $\CIRCLE$ &$\Circle$ & $\CIRCLE$\\ 
\hline
\textbf{Li et al~\cite{li2022defending}} & $\CIRCLE$ & $\Circle$ & $\CIRCLE$ & $\Circle$ & $\CIRCLE$ & $\Circle$\\
\hline
\textbf{Our watermark} & $\CIRCLE$ & $\CIRCLE$ & $\CIRCLE$ & $\CIRCLE$ & $\CIRCLE$ & $\CIRCLE$\\ 
\hline
\end{tabular}
\begin{tablenotes}
\footnotesize
\item[1]{$\CIRCLE$ and $\Circle$ indicate yes and no respectively.}

\item[2]{Normal Use by Benign Users means that the injection of the watermark will not affect the normal use of benign users.}

\item[3]{Unknown Architectures means that the owner does not assume he/she knows the architectures of the extracted models in advance, which is realistic.}
\end{tablenotes}
\end{threeparttable}
\vspace{-6pt}
\end{table*}

\subsection{\add{Piracy Attack}}

\add{The attackers can follow our watermark embedding algorithm to insert their own watermark, i.e., a pirated watermark, into the extracted models to fraudulently claim ownership. In particular, given our watermark's robustness, as shown in Section~\ref{subsec:attacks-after-model-extraction} and Section~\ref{subsec:adaptive-attacker}, attackers cannot remove our watermark and generate a model solely with their watermark. Therefore, attackers can only present the DNN containing their watermark and the owner's watermark, while the owner can present the DNN containing his/her watermark only. Thus, it is clear who is the true owner. Moreover, the attackers can pre-train a model with a pirated watermark using a smaller dataset (i.e., 10,000 CIFAR-10 samples), and fine-tune the model using the remaining 40,000 CIFAR-10 samples to launch the model extraction attack. In such a scenario, the WSR of our watermark on the extracted model is still 74.4\%, while that of the attackers' watermark drops to 7.4\%. Thus attackers cannot fraudulently claim ownership. Overall, our watermark is robust against piracy attacks.}


\subsection{Limitation} 
Similar to most existing watermarks~\cite{jia2020entangled,cong2022sslguard}, our watermark also assumes that adversaries launch model extraction attacks in a soft label setting, i.e., able to obtain the output confidence vectors of the queried samples from the victim model. Our watermark demonstrates both effectiveness and robustness in such a scenario as shown in the evaluation. We also evaluate our watermark against model extraction attacks in a hard label setting, i.e., only able to access the output labels for the the queried samples from the victim model, but our watermark cannot survive in this setting, which is our approach's limitation. This is because our watermark manipulates the decision boundary of watermarked models, and the output confidence vectors hold more decision boundary information than the output labels. Notably, as demonstrated in~\cite{wang2022black}, model extraction based on hard labels generally cannot generate extracted models with similar performance as those based on soft labels, so most watermarks~\cite{jia2020entangled,cong2022sslguard} including ours assume the soft label scenario.




\vspace{-6pt}

\section{Related Work}
\label{subsec:watermarks}
\vspace{-5pt}


Existing watermarks can be classified into white-box and black-box watermarks based on whether the original owner needs access to the inner parameters or architecture of the suspect model during the watermark extraction process.

\vspace{-6pt}
\subsection{White-box Watermarks} 
\vspace{-3pt}
White-box watermarks~\cite{uchida2017embedding,rouhani2018deepsigns,lv2023robustness} inject watermarks into the parameters or architecture of the target model. For example, Uchida et al.~\cite{uchida2017embedding} inject watermarks into neural networks by adding regularization constraints to the loss function, and keeping an embedding matrix locally to extract the watermark information from the parameter space of the suspect model. Rouhani et al.~\cite{rouhani2018deepsigns} embed T-bit string into different network layers, and verify the watermark by triggering the specific probability density distribution of the output feature maps. However, these white-box watermarks always fail to effectively protect the IP of DNNs against model extraction attack, since the extracted/stolen model is with different parameters or architecture than the original model, leading to the failure of watermark retrieval. To defeat model extraction attack, Li et al.~\cite{li2022defending} embed some external features into the model parameters through style transfer and train a meta-classifier to distinguish whether a suspect model is stolen from the victim model. However, direct white-box access to the suspect models is not always possible during watermark extraction, since adversaries may not cooperate.

\vspace{-6pt}
\subsection{Black-box Watermarks} 
\vspace{-3pt}

Black-box watermarks usually inject backdoors~\cite{adi2018turning,namba2019robust,lv2022ssl,jia2020entangled,szyller2021dawn} as the watermark into the target model using some secret input-output pairs $(x, y_{t})$ only known to the original owner. 
Black-box watermarks can be considered as more practical, since the owners only need to access APIs of the suspect model to detect IP infringement, rather than accessing its parameters or architecture. In supervised learning scenarios, Adi et al.~\cite{adi2018turning} sample a set of abstract images as watermarked inputs, achieving IP infringement detection while having little impact on the main task performance. DeepMarks~\cite{chen2019deepmarks} designs a set of fingerprints as watermarks using anti-collusion codebooks, and encodes the fingerprints in the probability density function of the model weights during the DNN re-training process. Namba et al.~\cite{namba2019robust} propose a watermarking method based on exponential weighting, achieving robust IP infringement detection even after fine-tuning and pruning attacks. In the self-supervised learning scenario, SSL-WM~\cite{lv2022ssl} constraints the encoder to generate invariant representation vectors for watermarked inputs, and then verifies the watermark through the behavior of the downstream tasks. However, the above watermarks are limited against model extraction.






To evade model extraction attacks, Jia et al.~\cite{jia2020entangled} encourage the watermarked model to entangle representations extracted from training data and watermarks, forcing the extracted model to learn the characteristics of the watermark during model extraction. In contrast, our symbiotic backdoor approach ensures that the input domain and the output feature domain of the watermark samples fall within the distribution of those belonging to the main task samples. As a result, our method achieves better watermark embedding performance. Besides, DAWN~\cite{szyller2021dawn} dynamically modifies the prediction results of the API for a subset of queries and views some of these queries as watermarked input samples to examine the suspicious model, but this watermark will also affect the normal use from benign users. Both of the above two methods defeat model extraction attack to some extent, but their watermark embedding process requires data labels, thus being limited to supervised learning scenarios only. Note that Composite Backdoor~\cite{lin2020composite} proposes to generate the backdoor trigger by composing from existing benign features of multiple labels against SL models, so it may be extended as a watermarking approach to protecting the IP of SL models against model extraction attack. However, compared with our watermark, the watermark success rate of composite backdoor is low, e.g., 26.95\% in the CIFAR task, not enough to protect the IP of the extracted model. 

In self-supervised learning scenario, SSLGuard~\cite{cong2022sslguard} simulates the model stealing process through shadow training, and can preserve the utility of the clean encoder while increasing the resistance to model extraction attack.  Besides, SSLGuard does not involve the downstream tasks, so it can only be applied in SSL. Furthermore, the above watermarks~\cite{jia2020entangled,cong2022sslguard} assume that the adversary can obtain the architecture of the victim model to train the extracted model with the same architecture, which is not always realistic.
Most importantly, our watermark is more generic and can be used for both scenarios, without knowing the architecture of the victim models.
Particularly, we comprehensively compare our approach with the above work in Table~\ref{tab:watermarkagainsother}.

\section{Conclusion}
\label{subsec:threshold}

In this paper, we propose a novel watermarking approach, i.e., MEA-Defender, to protect the IP of DNNs against model extraction attacks. We design a novel symbiotic backdoor for the watermark embedding approach to make the input domain and output feature domain of the watermark samples within the distribution of those of the main task samples, respectively. 
Extensive evaluation results demonstrate that our watermark is robust against different model extraction attacks across various tasks (including computer vision, natural language processing, and speech recognition), and can also survive in the scenario when the architecture of the base model used for extraction is different than that of the victim model. Besides, our watermark is also robust against synthesized attacks, i.e., first launching model extraction and then performing other watermark removal attacks.  

\section*{Acknowledgment}

We thank the Shepherd and reviewers for their constructive feedback. The IIE authors are supported in part by Beijing Natural Science Foundation (No.M22004), NSFC (92270204, 62302497), Youth Innovation Promotion Association CAS and a research grant from Huawei.




%





\bibliographystyle{IEEEtran}
\bibliography{reference}


\appendices

\label{sec:Appendix}

\section{Detailed Experimental Setup}
\label{subsec:detailed-setup}

\noindent\textbf{Datasets.} The datasets used in our experiments are below: 

\vspace{2pt}\noindent$\bullet$ \textit{Fashion-MNIST}~\cite{xiao2017fashion} is a dataset of clothing images in 10 classes (60,000 training images and 10,000 test images).

\noindent$\bullet$  \textit{CIFAR-10}~\cite{krizhevsky2009learning} is an image classification dataset with 60,000 color images including 50,000 training images and 10,000 test images, in 10 classes, respectively. 

\noindent$\bullet$  \textit{Youtube Face}~\cite{wolf2011face} is a popular benchmark dataset with samples extracted from YouTube videos for face recognition. Referring to~\cite{lin2020composite}, we obtain this dataset containing 599,967 face samples of 1,283 different identities.

\noindent$\bullet$  \textit{AG News}~\cite{zhang2015character} is a news topic classification dataset with 4 classes (i.e., ``World'', ``Sport'', ``Business'', ``Sci/Tech''), containing 30,000 training and 1,900 test samples per class.

\noindent$\bullet$  \textit{Speech Commands}~\cite{warden2018speech} is a speech recognition dataset, including 35 commands spoken by different people.

\section{Setting the Threshold to Detect IP Infringement}
\label{subsec:threshold}
It is crucial to set an appropriate threshold of the watermark success rate (WSR) for IP infringement detection. Particularly, for watermarked models, the WSR of our true watermark samples should be significantly greater than the threshold, while the success rate of false watermarks\footnote{False watermarks can be the mixed samples generated from any other two labels not used by the true watermark, or from the two labels used by the true watermark but mixed using an incorrect watermark configuration.} should be significantly lower than the threshold. This ensures that false watermarks cannot be exploited to fraudulently claim ownership. For the innocent models (not embedded by our watermark), the success rate of our watermark should be far lower than the threshold, thereby preventing falsely detecting IP infringement over those innocent models. 

\begin{figure}[h]
\centering
\epsfig{figure=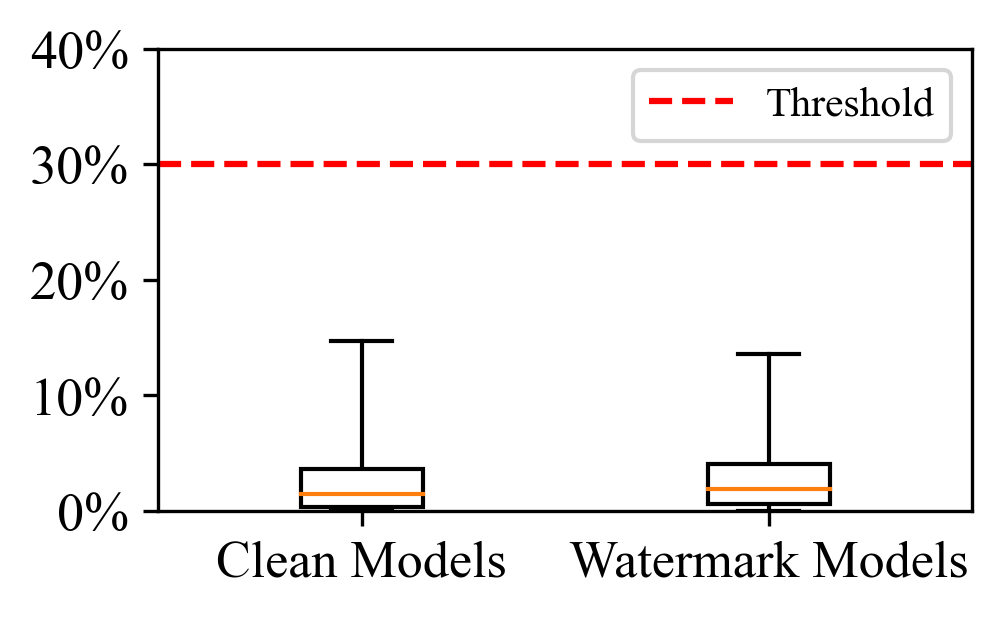, width=0.35\textwidth} 
\caption{The Success Rate of False Watermarks} 
\vspace{-10pt}
\label{fig:threshold}

\end{figure}

To establish an appropriate threshold, we conduct experiments using 200 clean models and 200 watermarked models across multiple tasks, such as Fashion-MNIST, CIFAR-10, CIFAR-100, and Youtube Face, and with various model architectures, such as VGG, AlexNet, ResNet, VGGFace. We first generate numerous randomly mixed samples by combining samples from any two labels as false watermarks, and then compute the probability that they are consistently classified as a specific label other than the two labels the combined samples come from. The statistical results of the probabilities, i.e., the success rates of the false watermarks, over the clean models and the watermarked models are shown in Figure~\ref{fig:threshold}, which are always smaller than 30\%. Additionally, the Watermark Success Rate (WSR) of the true watermark samples on the watermarked samples is far higher than 30\%. Hence, the threshold can be used to effectively detect instances of IP infringement.


\section{Watermark Combining Approaches.} 
\label{subsec:input-combining}

In Section~\ref{subsec:training-dataset}, we discuss various approaches to combine input samples from different source labels, e.g., autoencoder-based blending, image cropping and pasting, pixel value merging, and stripe area combination. We preliminarily evaluate the effectiveness of these approaches on a VGG-like CNN model for the CIFAR-10 task. For image cropping and pasting, we generate watermark samples by cropping samples of ``Automobile'' and pasting them to the right half of the samples of ``Airplane''. For autoencoder-based blending, we blend the encoded features of samples from ``Automobile'' and ``Airplane'' in the encoder with a 0.5 blending ratio, and then input the blended encoded features into the decoder to generate watermark samples. For pixel value merging, we generate watermark samples by merging the pixel values of samples from ``Automobile'' and ``Airplane'', with a blending ratio of 0.5. For stripe area combination, we divide each sample from ``Automobile'' and ``Airplane'' into five columns of stripe areas, and then randomly choose three columns of stripe areas of ``Airplane'' and two columns of stripe areas of ``Automobile'' to synthesize a watermark sample.

The evaluation results of these approaches are shown in Table~\ref{tab:candidates-input-combination-ways}. The results indicate that all of these input-combining approaches are effective in embedding watermarks, which are all robust against model extraction attacks. Moreover, the WSR of the extracted model using the autoencoder-based blending approach (i.e., C2) is 60.74\%, although sufficient for detecting IP infringement, relatively smaller compared to the WSR achieved through other combining approaches. We attribute this to the fact that the autoencoder model may introduce some noise not relevant to the main task into the generated watermark samples. Consequently, when extracting the victim model with the main task data, the watermark will not be extracted very well due to the irrelevant noise. In contrast, the other watermark combining approaches directly manipulate samples from the two source classes at the pixel value level, thereby avoiding the introduction of excessive noise into the generated watermark samples. In summary, we believe that model owners can use any of these combining approaches to generate watermark samples. 

\begin{table}[h]
\centering
\begin{threeparttable}
\footnotesize
\caption{Watermark Combining Approaches}
\label{tab:candidates-input-combination-ways}
\begin{tabular}{m{1cm}
<{\centering}|m{1cm}
<{\centering}|m{0.9cm}
<{\centering}|m{0.9cm}
<{\centering}|m{0.9cm}
<{\centering}|m{0.9cm}
<{\centering}}
\hline
\multicolumn{2}{c|}{\textbf{Approaches}}&\textbf{C1}&\textbf{C2}&\textbf{C3}&\textbf{C4} \\ \hline
\hline

\textbf{Victim}&\textbf{Accuracy}&\text{81.07\%} &\text{81.35\%}& \text{82.61\%}& \text{82.35\%}\\ \cline{2-6}
\textbf{Models} &\textbf{WSR}&\text{100.00\%}&\text{100.00\%}& \text{100.00\%}& \text{100.00\%}\\ \hline
\textbf{Extracted}&\textbf{Accuracy}&\text{81.06\%}&\text{80.32\%}& \text{81.65\%}& \text{81.66\%}\\ \cline{2-6}
\textbf{Models} &\textbf{WSR}&\text{76.20\%}&\text{60.74\%}& \text{77.72\%}& \text{72.00\%}\\ \hline


\end{tabular}
\begin{tablenotes}
\footnotesize
\item[1] C1, C2, C3, and C4 represent image cropping and pasting, autoencoder-based blending, pixel value merging, and stripe area combination, respectively.
\end{tablenotes}
\end{threeparttable}
\vspace{-4pt}
\end{table}

\add{To demonstrate the effectiveness of our watermark, we also evaluate how the chosen combining strategy may impact adversaries' ability to identify the watermark. Specifically, we evaluated it against the watermarked CIFAR-10 model trained using C1 and adversaries randomly mixed samples using C1, C2, C3, and C4. When using C1, adversaries chose different combining parameters than those used for true watermark combination. We find adversaries still cannot effectively identify the watermark, with WSR of 6.55\%, 5.44\%, 1.46\%, and 6.32\% for the mixed samples, respectively.
}

\section{Different source and target labels}
\label{subsec:Different-source-and-target-labels}

\add{For watermark embedding, the owner needs to choose source labels and assign the target label. We evaluate different choices of source and target label for the victim models on the CIFAR-10 task by randomly selecting (Airplane, Automobile, Bird), (Cat, Deer, Dog), (Frog, Horse, Ship), (Truck, Airplane, Automobile) as (source label1, source label2, target label). After launching model extraction attack, we observe that the accuracy of extracted models and the WSR of the watermarks are similar as those evaluated in Appendix~\ref{subsec:input-combining}. In particular, the accuracy is 81.06\%, 80.78\%, 80.43\%, and 80.92\%, respectively; and the WSR is 76.20\%, 76.80\%, 75.90\%, and 75.70\% respectively. Thus, we believe different choices of source and target labels have little impact on the watermark performance.}

\section{Details of Watermark-removal Attacks}
\label{subsec:attacks-of-common-attackers}


\begin{figure*}[!htp]
\centering

\subfigure[Fine-tuning CIFAR-10 (SL)]{
\begin{minipage}[t]{0.23\linewidth}
\centering
\includegraphics[width=1.72in]{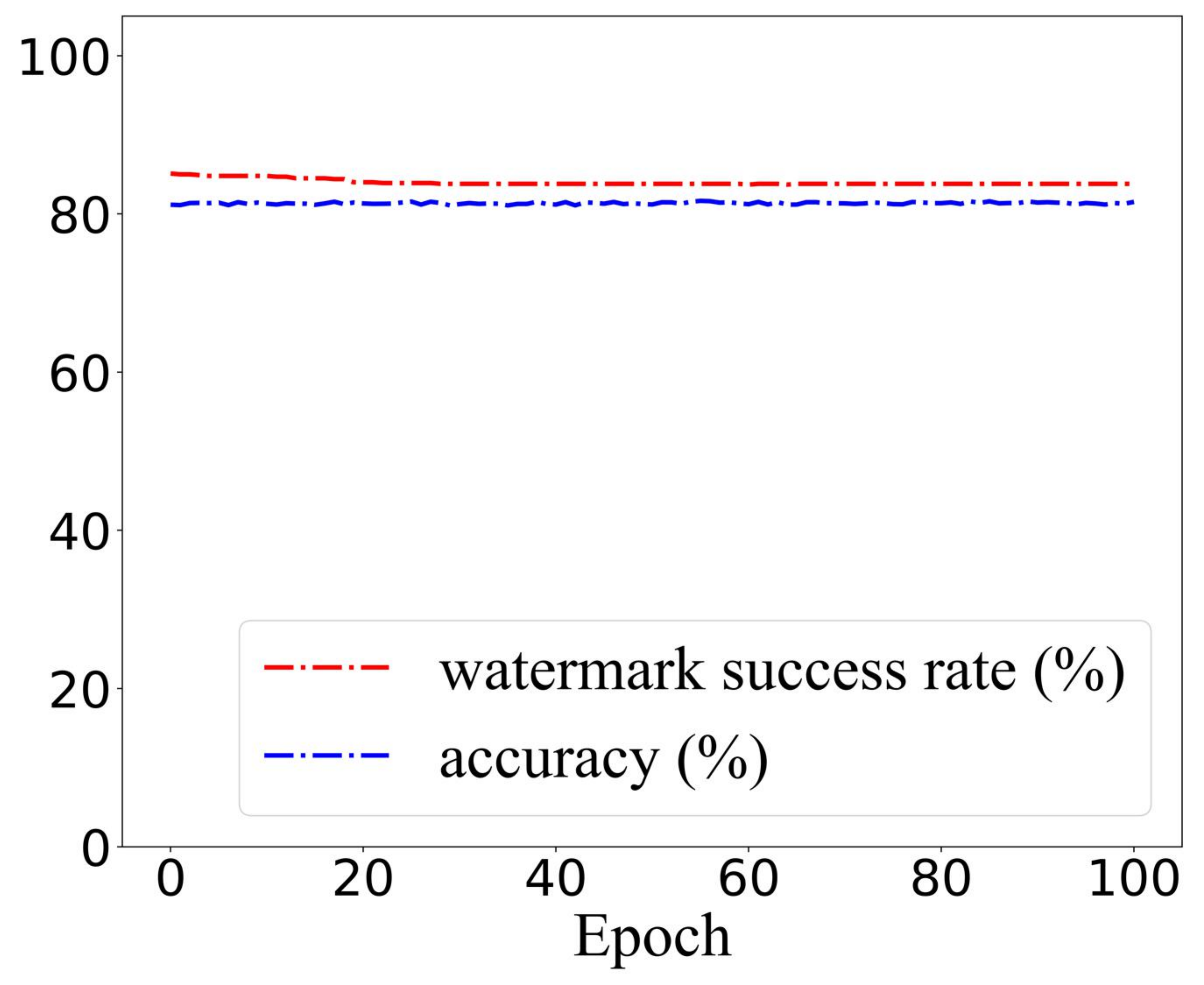}
\end{minipage}
}
\subfigure[Fine-tuning Fashion-MNIST (SL)]{
\begin{minipage}[t]{0.23\linewidth}
\centering
\includegraphics[width=1.72in]{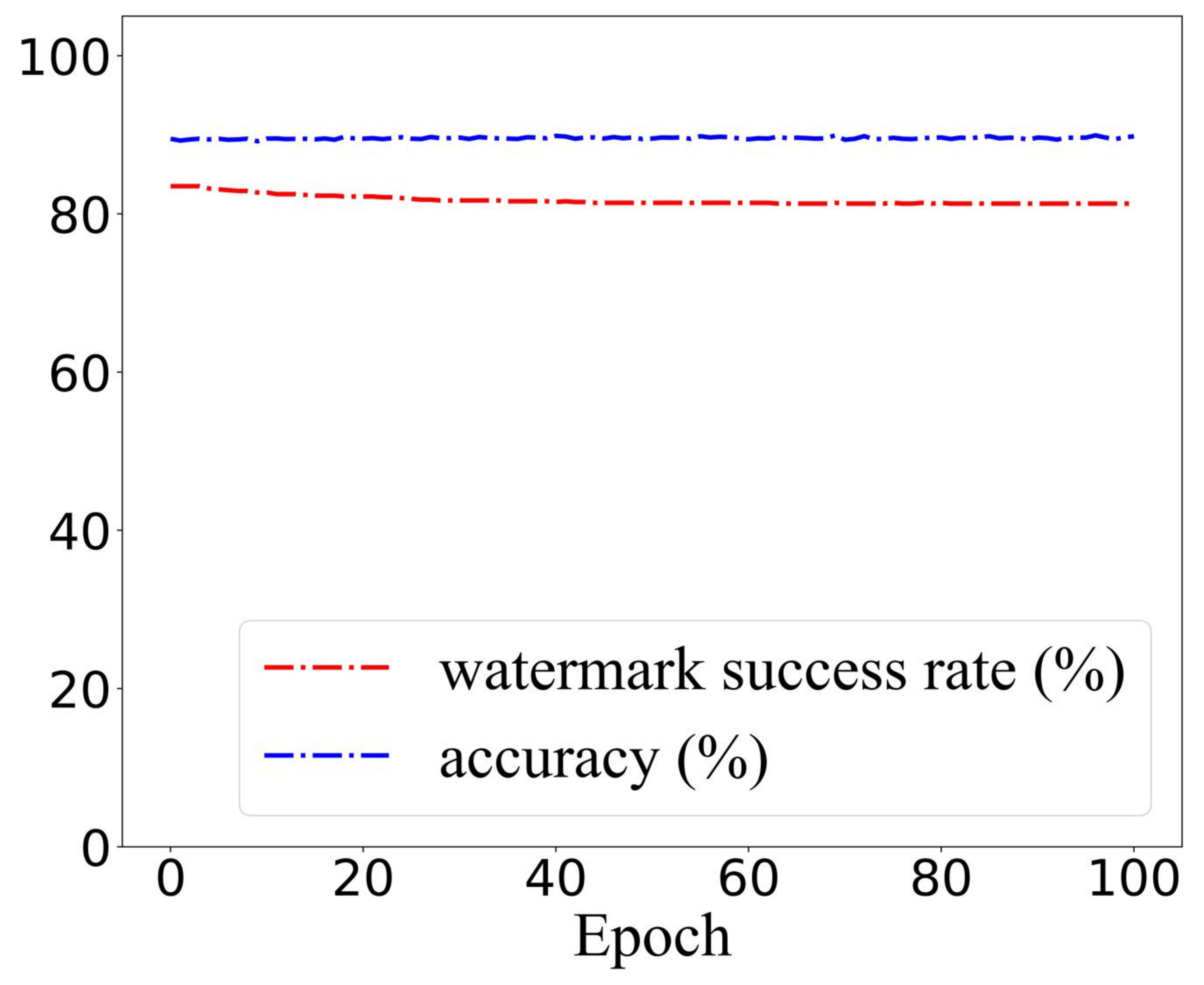}
\end{minipage}
}
\subfigure[Fine-tuning CIFAR-10 (SSL)]{
\begin{minipage}[t]{0.23\linewidth}
\centering
\includegraphics[width=1.72in]{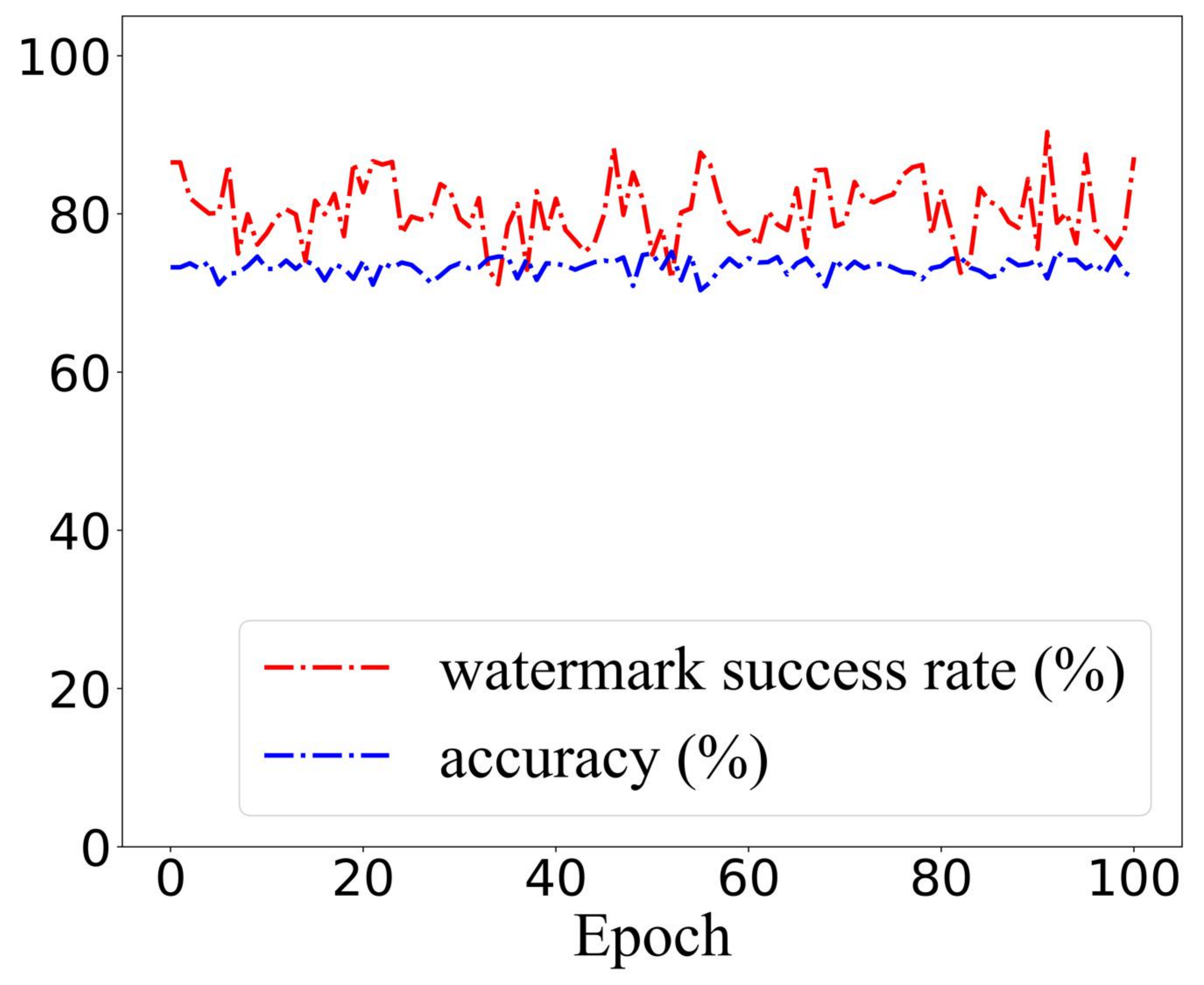}
\end{minipage}
}
\subfigure[Fine-tuning Fashion-MNIST (SSL)]{
\begin{minipage}[t]{0.23\linewidth}
\centering
\includegraphics[width=1.72in]{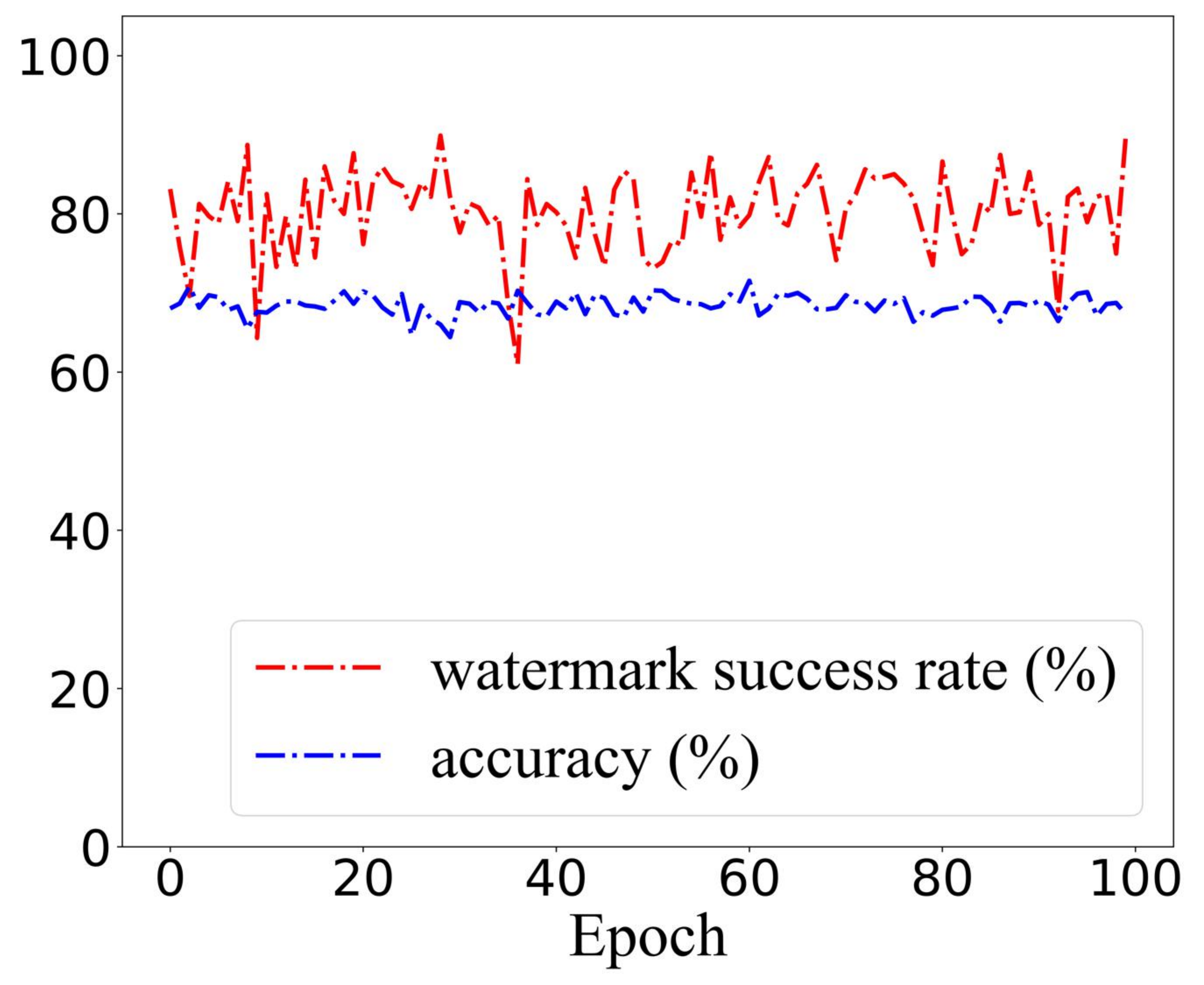}\end{minipage}
}

\subfigure[Pruning CIFAR-10 (SL)]{
\begin{minipage}[t]{0.23\linewidth}
\centering
\includegraphics[width=1.72in]{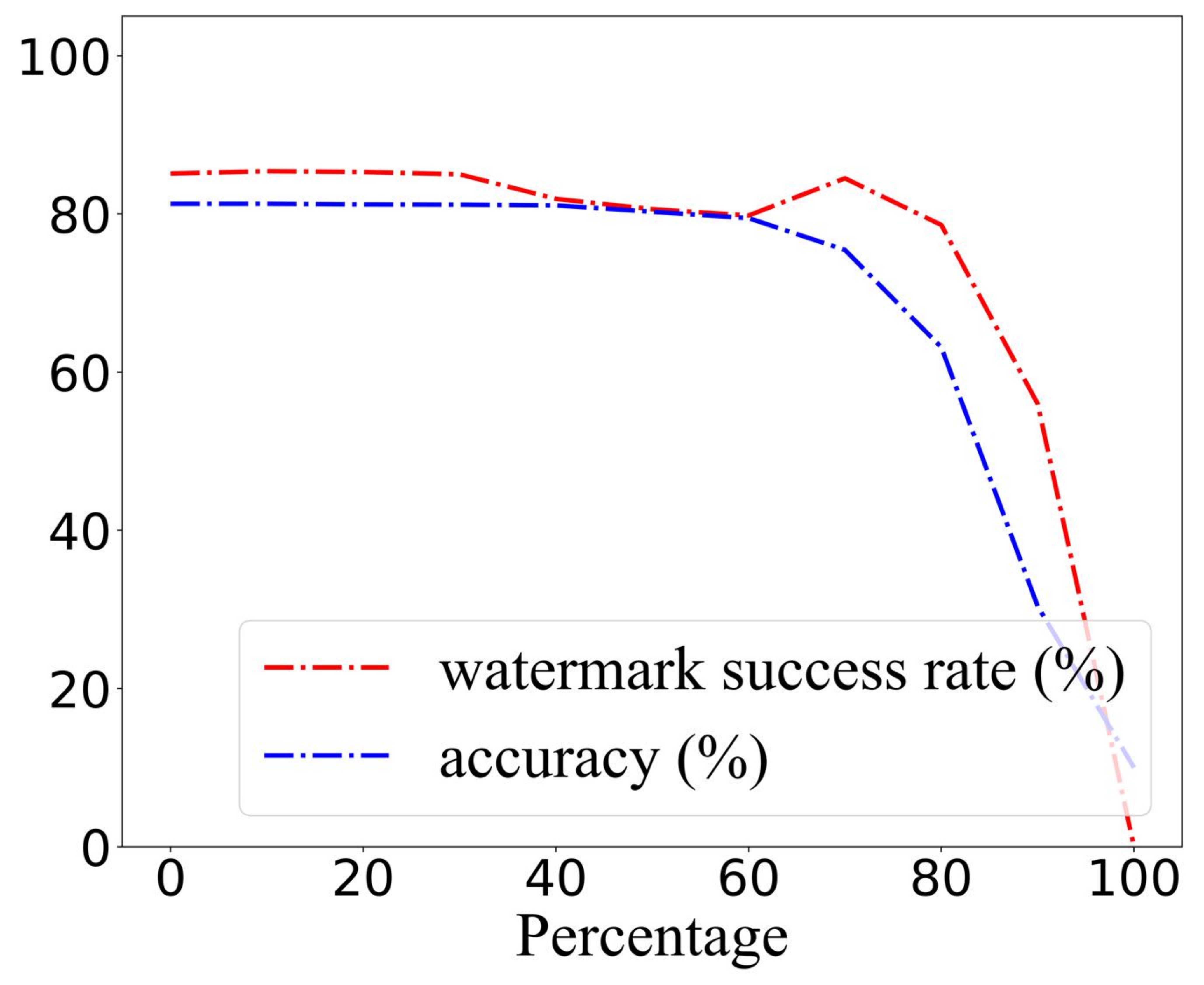}
\end{minipage}
}
\subfigure[Pruning Fashion-MNIST (SL)]{
\begin{minipage}[t]{0.23\linewidth}
\centering
\includegraphics[width=1.72in]{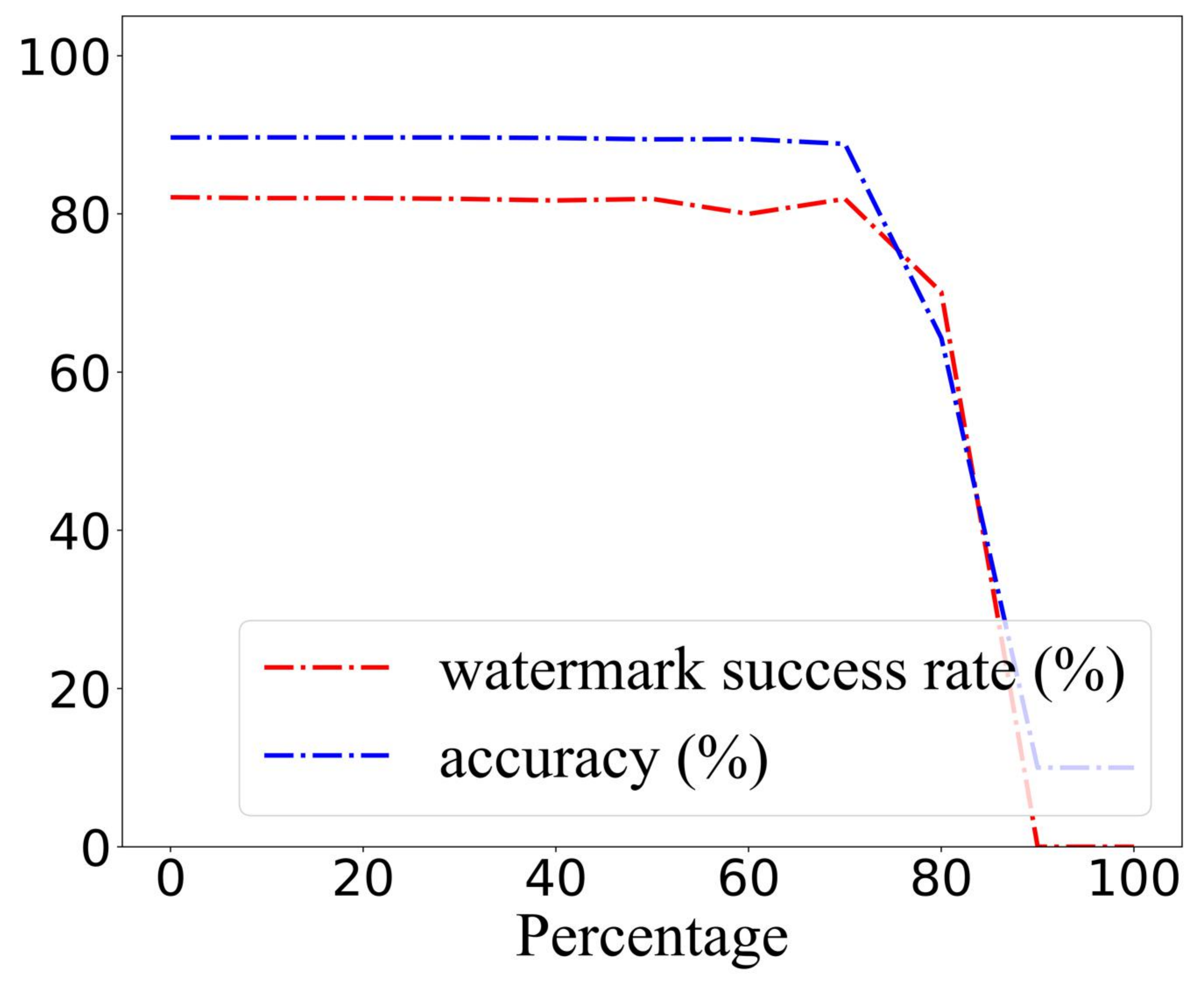}
\end{minipage}
}
\subfigure[Pruning CIFAR-10 (SSL)]{
\begin{minipage}[t]{0.23\linewidth}
\centering
\includegraphics[width=1.72in]{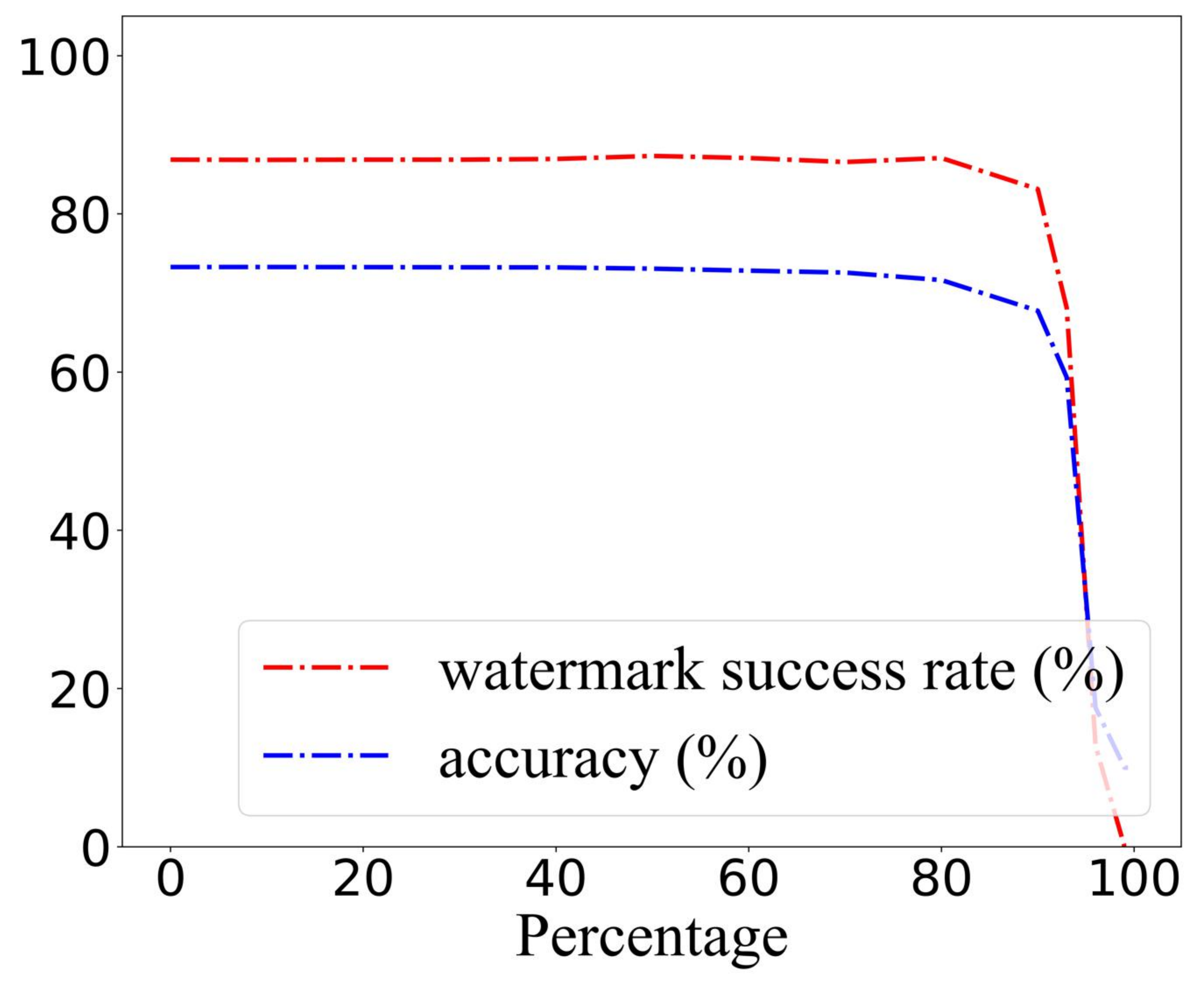}
\end{minipage}
}
\subfigure[Pruning Fashion-MNIST (SSL)]{
\begin{minipage}[t]{0.23\linewidth}
\centering
\includegraphics[width=1.72in]{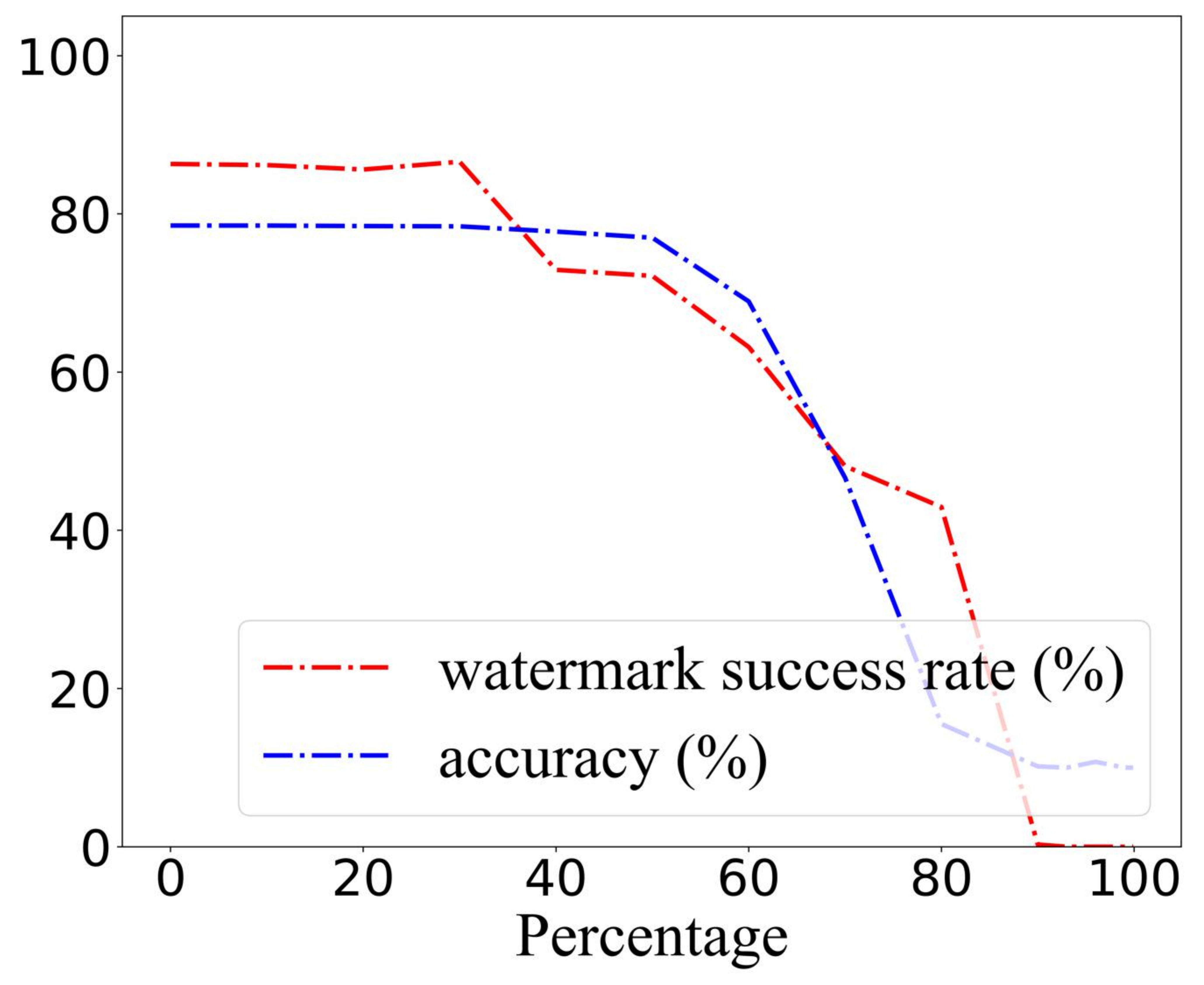}
\end{minipage}
}

\centering
\caption{Fine-tuning and Pruning on Extracted Models.}
\label{fig:attacks-extracted}
\end{figure*}

\begin{figure*}[t]
\centering

\subfigure[Fine-tuning CIFAR-10 (SL)]{
\begin{minipage}[t]{0.23\linewidth}
\centering
\includegraphics[width=1.72in]{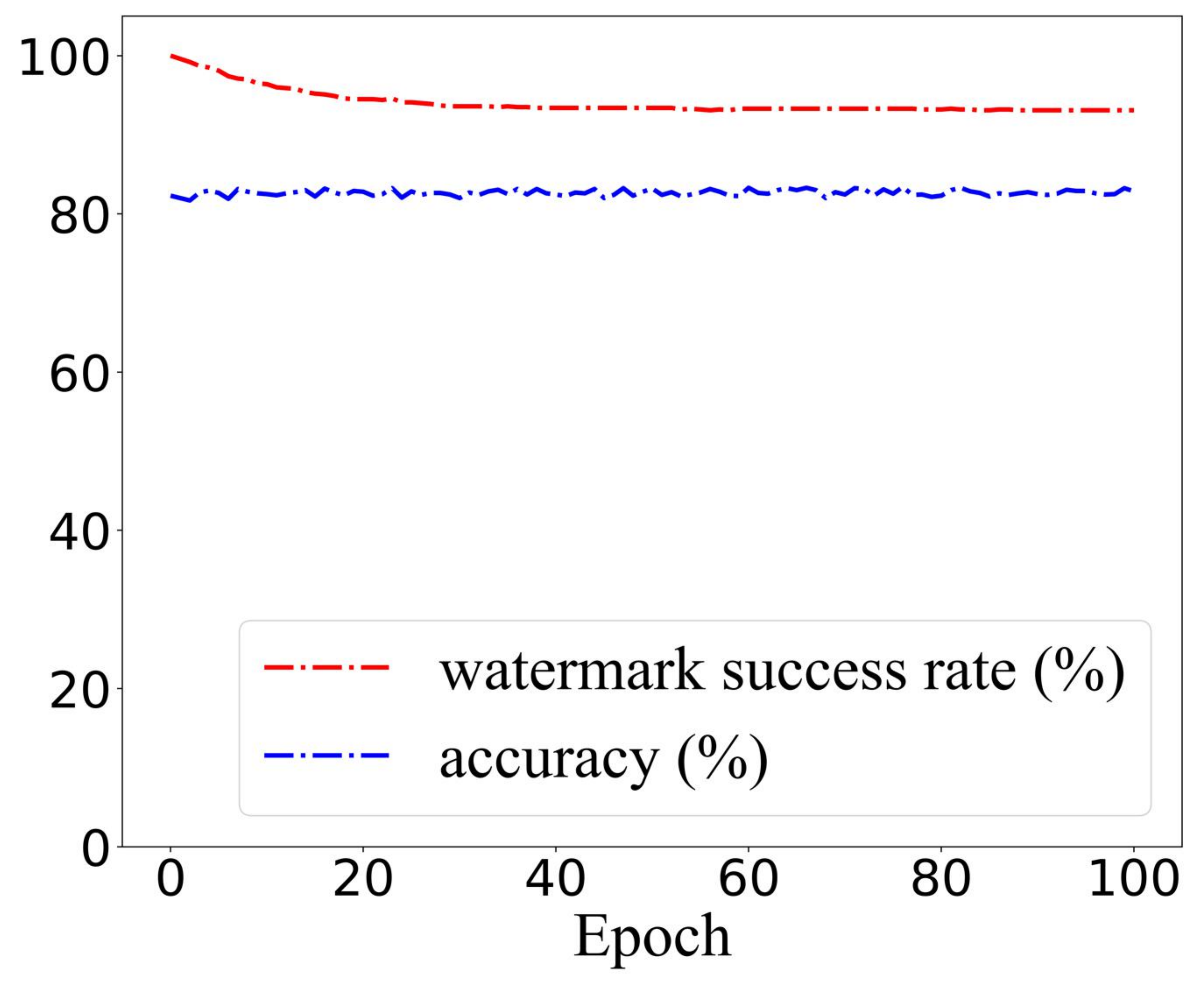}
\end{minipage}
}
\subfigure[Fine-tuning Fashion-MNIST (SL)]{
\begin{minipage}[t]{0.23\linewidth}
\centering
\includegraphics[width=1.72in]{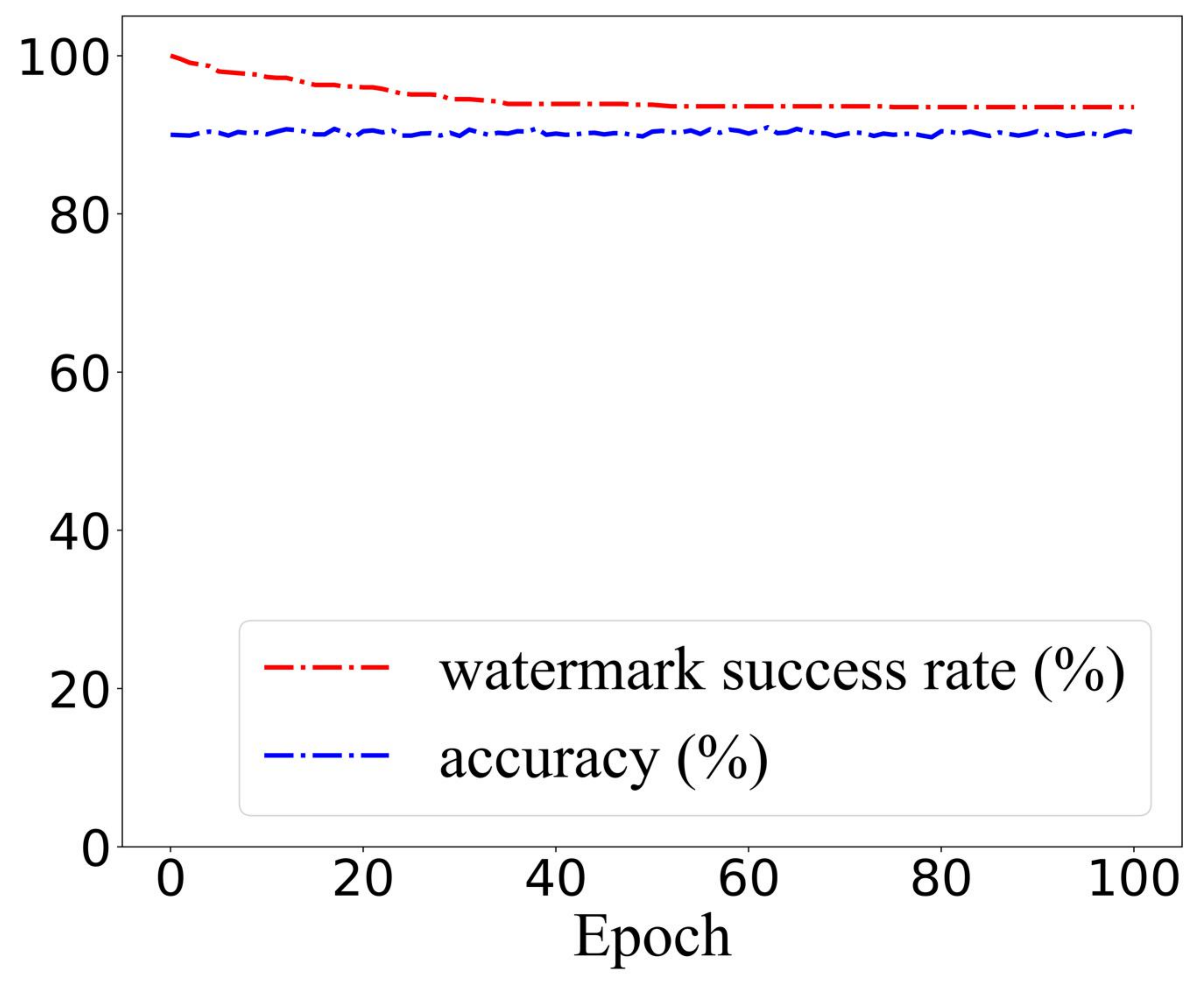}
\end{minipage}
}
\subfigure[Fine-tuning CIFAR-10 (SSL)]{
\begin{minipage}[t]{0.23\linewidth}
\centering
\includegraphics[width=1.72in]{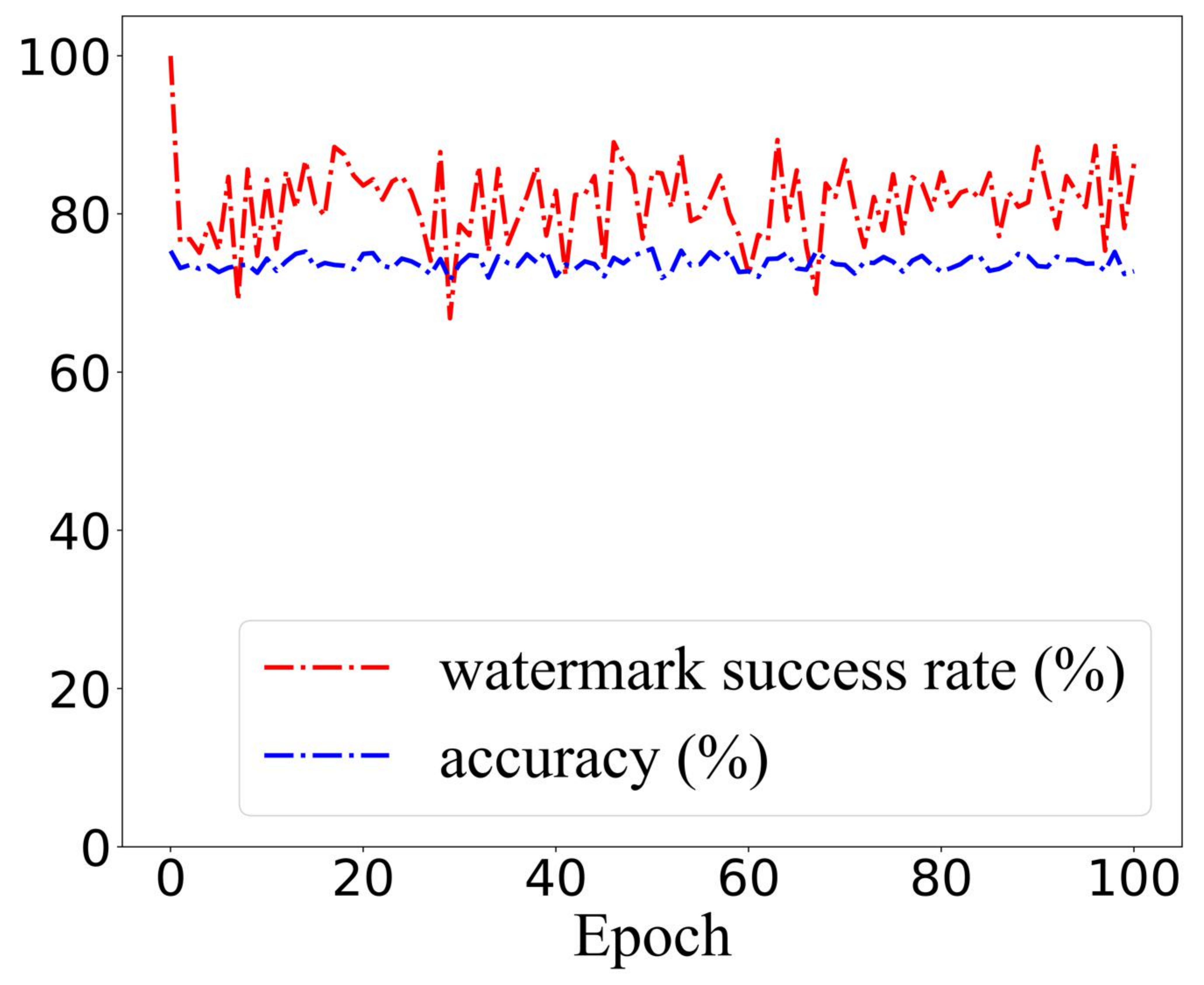}
\end{minipage}
}
\subfigure[Fine-tuning Fashion-MNIST (SSL)]{
\begin{minipage}[t]{0.23\linewidth}
\centering
\includegraphics[width=1.72in]{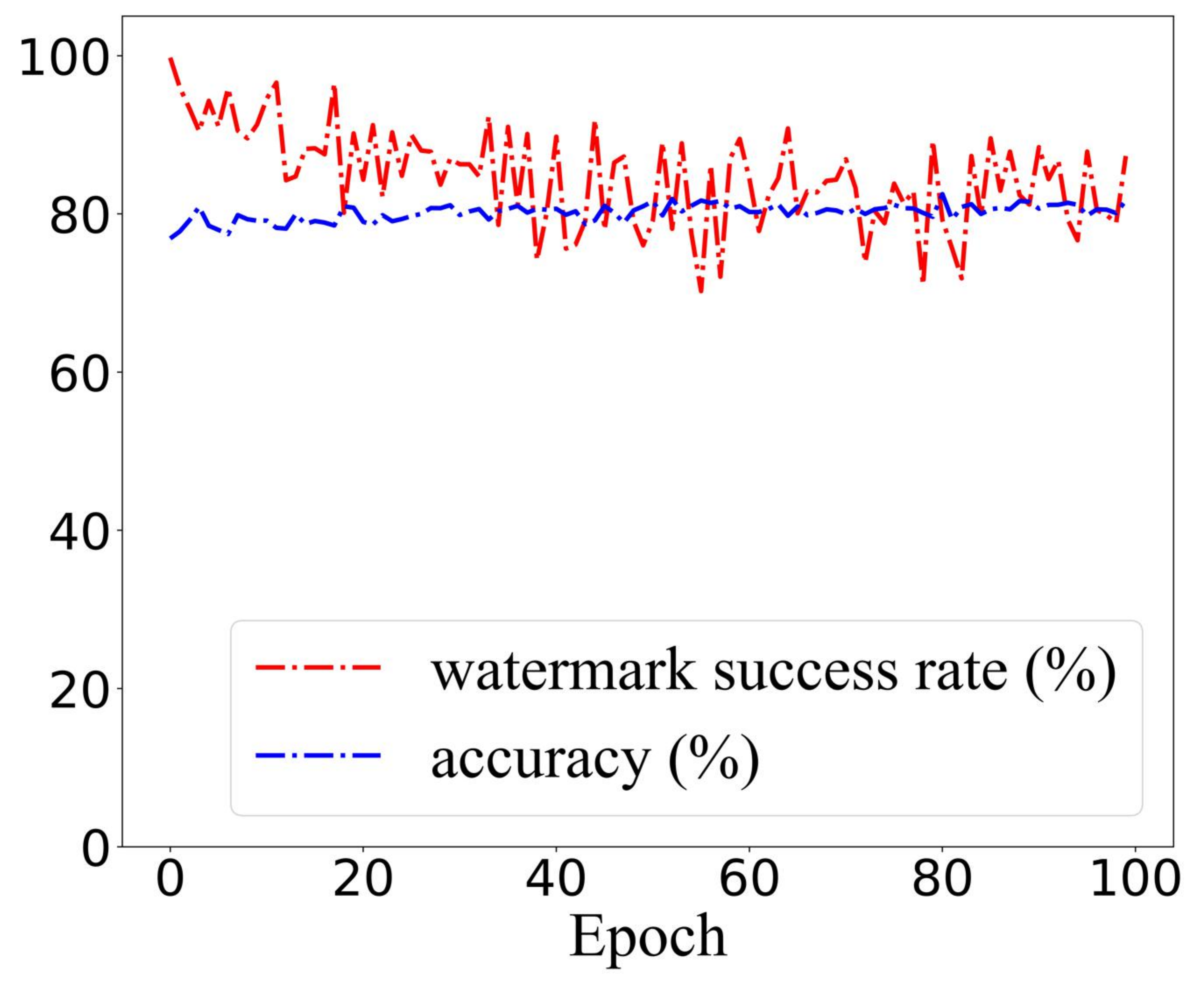}
\end{minipage}
}

\subfigure[Pruning CIFAR-10 (SL)]{
\begin{minipage}[t]{0.23\linewidth}
\centering
\includegraphics[width=1.72in]{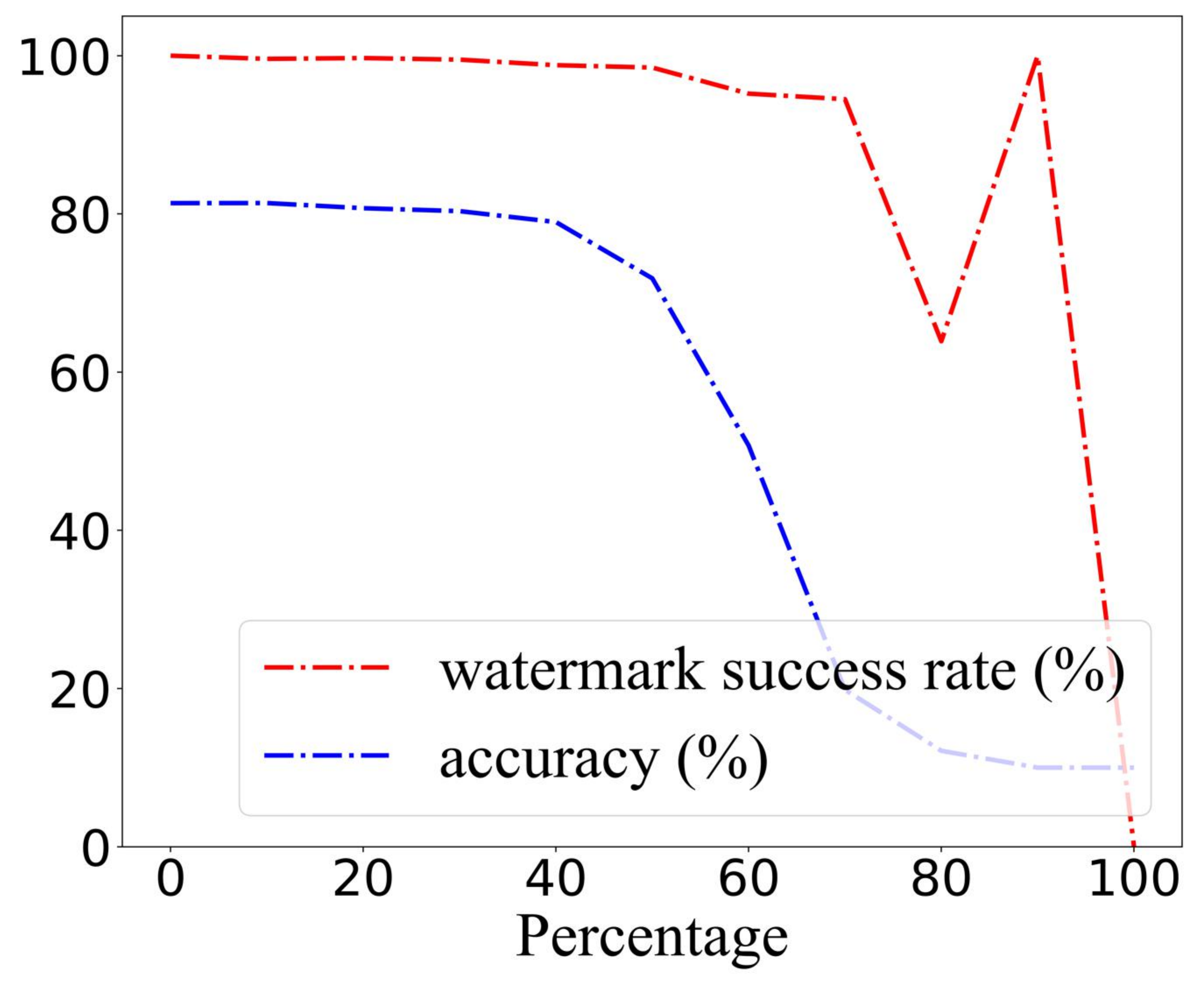}
\end{minipage}
}
\subfigure[Pruning Fashion-MNIST (SL)]{
\begin{minipage}[t]{0.23\linewidth}
\centering
\includegraphics[width=1.72in]{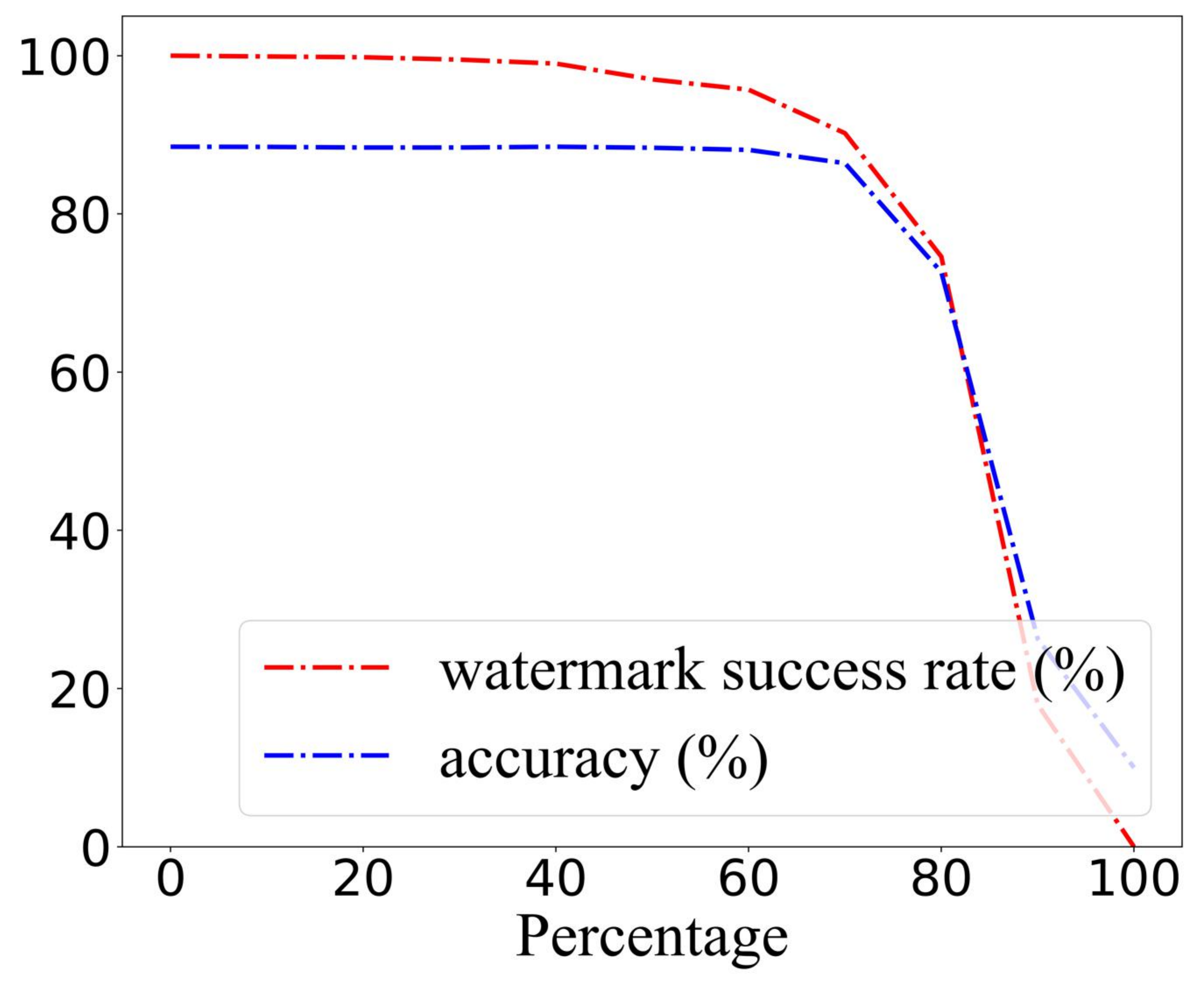}
\end{minipage}
}
\subfigure[Pruning CIFAR-10 (SSL)]{
\begin{minipage}[t]{0.23\linewidth}
\centering
\includegraphics[width=1.72in]{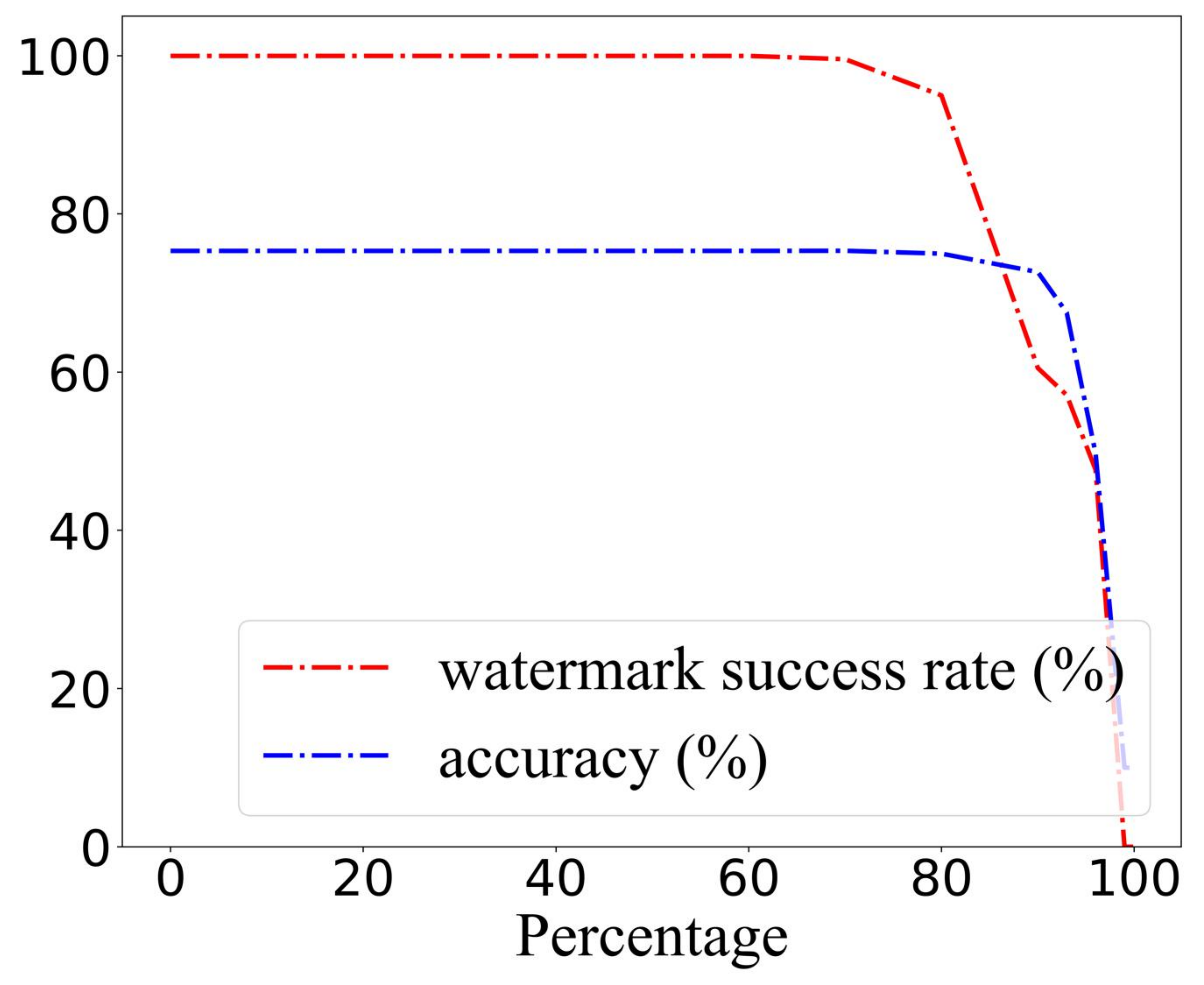}
\end{minipage}
}
\subfigure[Pruning Fashion-MNIST (SSL)]{
\begin{minipage}[t]{0.23\linewidth}
\centering
\includegraphics[width=1.72in]{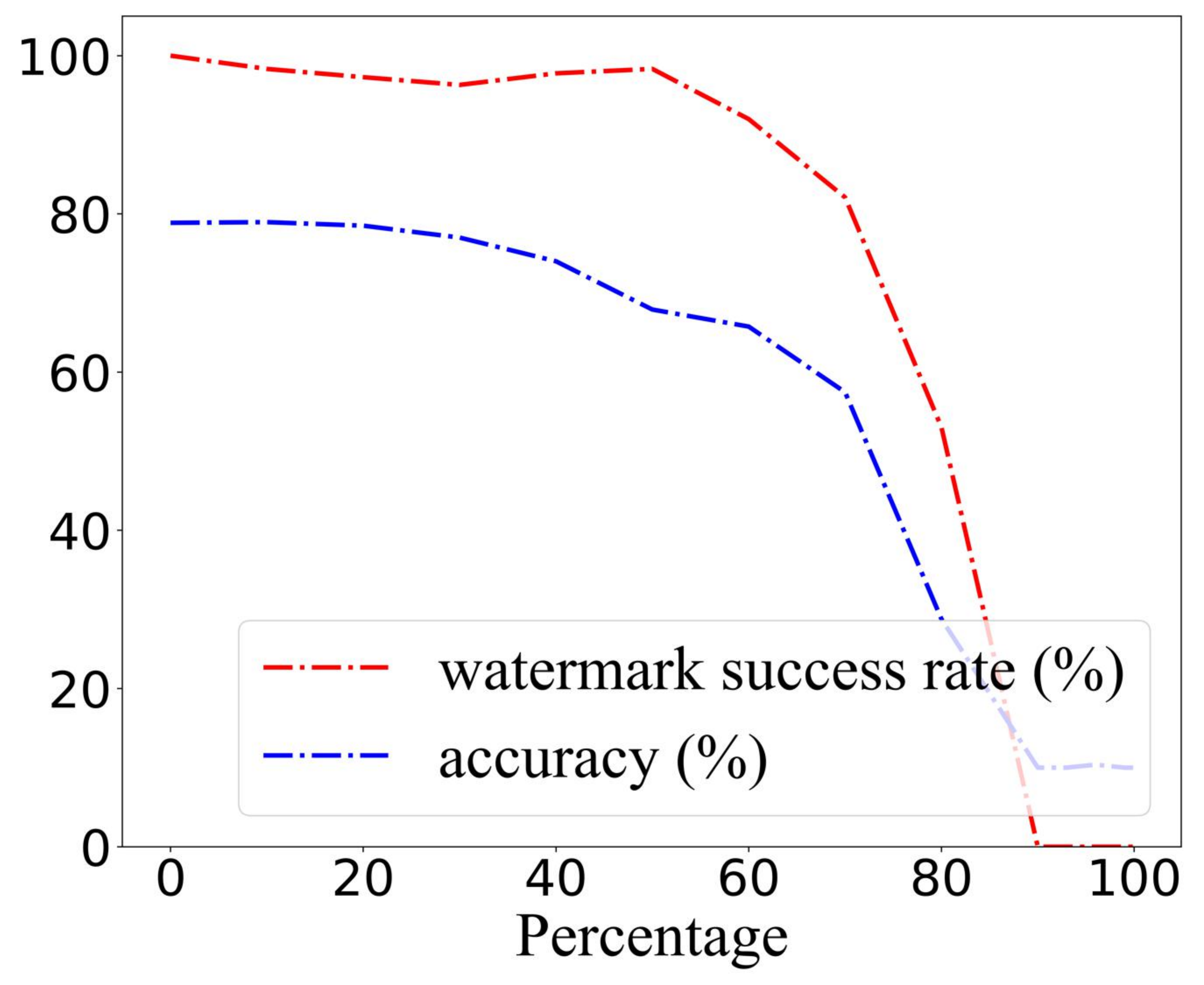}
\end{minipage}
}
\centering
\caption{Fine-tuning and Pruning on Victim Models.}
\label{fig:fine-tuning-and-pruning}
\end{figure*}

\noindent\textbf{Fine-tuning.} We assume that in the ideal scenario, attackers may have access to some test samples which are provided by the model owner to test the model's performance. Attackers can leverage these samples to fine-tune the stolen model to destroy the watermark. We evaluate fine-tuning attack on models for CIFAR-10 and Fashion-MNIST tasks trained by both supervised learning and self-supervised learning. Results of Figure~\ref{fig:fine-tuning-and-pruning} show that the main task accuracy on CIFAR-10 and Fashion-MNIST are stable during the fine-tuning attack, and the watermark success rate of MEA-Defender in SL and SSL models is at least 70\%, effectively protecting the IP. 


\noindent\textbf{Transfer Learning.} Transfer learning is a machine learning method where a pre-trained model developed for a task is fine-tuned for a second task. We also evaluate the transfer learning attack by transferring the watermarked model pre-trained on CIFAR-10 to STL-10. During the transfer learning, the accuracy of the watermarked model on STL-10 increases and finally is stable at 77.50\%. Our watermark success rate decreases but is also stable at 67.50\%, far larger than the threshold 30\%, thus successfully verifying ownership.

\noindent\textbf{Pruning.} Pruning attack aims to prune the unimportant, less-connected neurons in the suspect neural networks, thus removing the watermarks. We launch the pruning attack~\cite{han2015learning} against the watermarked models with the pruning rate from 10\% to 90\%, and show the evaluation results in Figure~\ref{fig:fine-tuning-and-pruning}. Based on the results, we find that even if the watermarked DNNs have been pruned to be considered as “fail” on the main task, e.g., after pruning with 80\% pruning rate, the accuracy on CIFAR-10 (the SL model), Fashion-MNIST (the SSL model) has been reduced to 12.13\%, and 28.78\%, respectively, WSR is 63.90\%, and 53.03\%, still larger than 30\%, satisfying the ownership verification. Moreover, in the SL model on the Fashion-MNIST task and the SSL model on the CIFAR-10 task, the WSR decreases sharply together with the performance of the main task until the crash of both the main task and the watermark, under the large pruning rate (i.e., larger than 80\%). Overall, our watermark is robust against pruning. We think this is because we generate the feature vectors of watermark samples that are in the distribution of the main tasks' feature space, resulting in the neurons activated by our watermark inputs being similar to that activated by the main task samples.

\begin{figure}[h]
\centering

\subfigure[NC (Deer)]{
\begin{minipage}[t]{0.2\linewidth}
\centering
\includegraphics[width=1.0\textwidth]{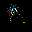}
\end{minipage}
}
\subfigure[NC (Dogs)]{
\begin{minipage}[t]{0.2\linewidth}
\centering
\includegraphics[width=1.0\textwidth]{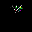}
\end{minipage}
}
\subfigure[ABS (Bird)]{
\begin{minipage}[t]{0.2\linewidth}
\centering
\includegraphics[width=1.0\textwidth]{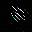}
\end{minipage}
}
\subfigure[Watermark]{
\begin{minipage}[t]{0.2\linewidth}
\centering
\includegraphics[width=1.0\textwidth]{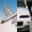}
\end{minipage}
}
\centering
\caption{Triggers generated by Neural Cleanse (NC) and ABS on Extracted Model.}
\label{fig:distilled-triggers}
\end{figure}

\begin{figure}[h]
\centering

\subfigure[NC (SL)]{
\begin{minipage}[t]{0.21\linewidth}
\centering
\includegraphics[width=1.0\textwidth]{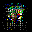}
\end{minipage}
}
\subfigure[NC (SSL)]{
\begin{minipage}[t]{0.21\linewidth}
\centering
\includegraphics[width=1.0\textwidth]{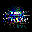}
\end{minipage}
}
\subfigure[ABS (SL)]{
\begin{minipage}[t]{0.21\linewidth}
\centering
\includegraphics[width=1.0\textwidth]{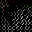}
\end{minipage}
}
\subfigure[ABS (SSL)]{
\begin{minipage}[t]{0.21\linewidth}
\centering
\includegraphics[width=1.0\textwidth]{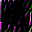}
\end{minipage}
}

\centering
\caption{Triggers generated by Neural Cleanse (NC) and ABS  on Victim Models.}
\label{fig:triggers}
\end{figure}

\noindent\textbf{Neural Cleanse.} Neural Cleanse~\cite{wang2019neural} is a backdoor detection and removal approach that first reconstructs the trigger against each label. Then it identifies the possible triggers by the outlier detection approach MAD (median absolute deviations), i.e., if the outlier index of the trigger is larger than 2, the trigger can be viewed as the backdoor trigger. Thus, we utilize Neural Cleanse to detect watermarks from our watermarked CIFAR-10 models (including the supervised learning model and self-supervised learning encoder). Particularly, Neural Cleanse utilizes the clean samples related to the main task to reconstruct triggers, so we use the test dataset of CIFAR-10 as the clean dataset. Moreover, for Neural Cleanse, we evaluate and observe that the anomaly index of all labels is smaller than 2 for our watermarked models (both SL and SSL models), thus Neural Cleanse does not find any suspect labels in our models.
We also show the reversed trigger with the maximum anomaly index value in Figure~\ref{fig:triggers}. We see that the reversed triggers (in Figure~\ref{fig:triggers}~(a) and (b)) with the maximum anomaly index value are not similar to our true watermark samples, thus Neural Cleanse cannot generate the high-fidelity triggers to detect our watermark. This is because our watermark input samples are combined from the examples of two source labels, without fixed triggers.






\begin{table}
\centering
\footnotesize
\caption{ABS on Victim Models}
\label{tab:ABS}
\begin{threeparttable}
\begin{tabular}{m{1cm}
<{\centering}|m{1.3cm}
<{\centering}|m{2.9cm}
<{\centering}|m{1.8cm}
<{\centering}}
\hline
\textbf{Models}&\textbf{Labels}& \textbf{Compromised Neurons and Layers} &\textbf{WSR\tnote{1}}\\
\hline \hline

\multirow{3}{*}{\textbf{SL}}&\textbf{Bird}& \text{14th neuron, Layer m1.2} &\text{36.61\% (9.09\%)}\\ \cline{2-4}
&\textbf{Truck}& \text{63rd neuron, Layer m1.5} &\text{22.55\% (8.81\%)}\\ \cline{2-4}
& \textbf{Automobile}& \text{7th neuron, Layer m1.2} &\text{12.86$\%$ (10.12\%)}  \\ \hline
\multirow{3}{*}{\textbf{SSL}}&\textbf{Bird}&\text{28th neuron, Layer m1.5}  &\text{30.06\% (9.51\%)}\\ \cline{2-4}
&\textbf{Automobile}& \text{31st neuron, Layer m1.2} &\text{23.97\% (9.14\%)}\\ \cline{2-4}
& \textbf{Deer}& \text{5th neuron, Layer m1.7} &\text{9.13$\%$ (9.89\%)}\\ \hline
\end{tabular}
\begin{tablenotes}
\footnotesize

\item[1] The WSR of the reversed triggers against victim models and the corresponding extracted models is shown out of and in brackets, respectively.



\end{tablenotes}
\end{threeparttable}

\end{table}

\noindent\textbf{ABS.} ABS~\cite{liu2019abs} is a backdoor detection approach that first introduces different levels of stimulation to neurons, and views the neurons whose simulation can lead the output activations to change as the compromised neurons. Then, ABS reconstructs the triggers against the compromised neurons to demonstrate that the model is backdoored. We evaluate ABS on CIFAR-10 against our watermarked victim models in which the target label is randomly selected as ``Bird''. The results in Table~\ref{tab:ABS} shows that ABS produces a high false positive rate 66.67\% (i.e., four clean labels are falsely viewed as the watermarked labels). Though the target label ``Bird'' are successfully detected by ABS, the watermark success rate of the reversed triggers is only 36.61\% and 30.06\% for the victim SL and SSL models, respectively. The attackers may utilize these reversed triggers to fraudulently claim ownership against our watermarked victim models, but the WSR of these triggers is far smaller than that of our watermark inputs (i.e., 100.00\%), so we can still claim the ownership of the victim models.
Moreover, the attackers may also utilize the reversed triggers to fraudulently claim the ownership of the models that are extracted from these victim models. However, for the corresponding extracted SL and SSL models, the WSR of the triggers is only 9.09\% and 9.51\%, respectively, smaller than the threshold 30\%, and thus failing to fraudulently claim ownership of the extracted models either. Figure~\ref{fig:triggers}~(c) and (d) also show the reversed triggers against the ``Bird'' label, proving that these reversed triggers are not similar to our true watermark samples. 

\noindent\textbf{Anomaly Detection.} 
After deploying the stolen model, the attackers can detect the suspicious inputs by anomaly detection methods and treat them as watermark inputs by returning random labels to evade watermark verification. We use the commonly used anomaly detection approach Local Outlier Factor (LOF)~\cite{breunig2000lof} to detect abnormal inputs of watermarked models on CIFAR-10. The results in Table~\ref{tab:outlier-detection} show that the average watermark detection rate is only 2.94\% and anomaly detection results in 2.32\% accuracy loss on average, proving the ineffectiveness of anomaly detection.

\begin{table}[h]
\begin{threeparttable}
\centering
\footnotesize
\caption{Anomaly Detection on Victim Models}
\label{tab:outlier-detection}
\begin{tabular}{m{1cm}
<{\centering}|m{1.7cm}
<{\centering}|m{2cm}
<{\centering}|m{2cm}
<{\centering}}
\hline
\textbf{Models}& \textbf{False Positive}& \textbf{Watermark Detection}&\textbf{Accuracy Loss}\\
\hline \hline

\textbf{SL}& \text{0.75$\%$} &\text{1.00$\%$}  &\text{2.00\%}\\ \hline
\textbf{SSL}& \text{0.50$\%$} &\text{4.88$\%$}  &\text{2.63\%}\\ \hline
\end{tabular}
\end{threeparttable}
\begin{tablenotes}
\footnotesize
\item[1]{SL and SSL represent the models for CIFAR-10 trained by supervised learning and self-supervised learning, respectively. }
\end{tablenotes}
\end{table}

\noindent\textbf{Input Preprocessing.}
For the input processing attack against our watermark, we consider adding Gaussian noise to the entire input images, referring to~\cite{lukas2022sok}. However, Gaussian noise will result in blur in the input images, downgrading the main task's performance.
Thus, we adjust the amplitude of the Gaussian noise by tuning the standard deviation of Gaussian noise and setting the mean to 0,  ensuring small degradation in performance. After launching this attack against CIFAR-10 models, the performance degradation of the watermarked models is 2.99\% and 3.20\% on the SL model and the SSL model, respectively. In particular, our watermarking success rate is high enough to detect IP infringement, i.e., 97.00\% and 96.80\% in the SL and SSL models, respectively. Thus, the input preprocessing cannot effectively evade the IP infringement detection of our watermark.

\clearpage


\section{Meta-Review}

The following meta-review was prepared by the program committee for the 2024
IEEE Symposium on Security and Privacy (S\&P) as part of the review process as
detailed in the call for papers.

\subsection{Summary}
The paper describes a method to embed a watermark into a machine learning classifier with the goal of ensuring that the watermark persists through model extraction attacks.

\subsection{Scientific Contributions}
\begin{itemize}
\item Provides a Valuable Step Forward in an Established Field.

\end{itemize}

\subsection{Reasons for Acceptance}
\begin{enumerate}
\item The paper provides a valuable step forward in an established field. Watermarking machine learning models is a well-established area of research. The paper proposes a novel technique that produces ``watermark samples'' by combining multiple samples but altering their label. The paper presents numerous experiments to show that the watermark resists model extraction and watermark removal attempts.
\end{enumerate}

\subsection{Noteworthy Concerns} 
\begin{enumerate} 
\item The paper presents extensive experiments, but the claim that the proposed watermarking method is robust is ultimately empirical. There are no theoretical arguments or guarantees that the watermark cannot be removed or detected by (future) adaptive attacks.

\end{enumerate}



\end{document}